\documentclass{pasj00}
\draft

\RelativeFigurePath{fig}

\begin{document}
\SetRunningHead{M. Kokubun et al.}
{In-Orbit Performance of the Hard X-ray Detector}
\Received{2006/09/06}
\Accepted{2006/10/15}

\title{In-Orbit Performance of the Hard X-ray Detector on board Suzaku}

%
\author{%
  Motohide \textsc{Kokubun}\altaffilmark{1},
  Kazuo \textsc{Makishima}\altaffilmark{1,2},
  Tadayuki \textsc{Takahashi}\altaffilmark{3,1},
  Toshio \textsc{Murakami}\altaffilmark{4},\\
  Makoto \textsc{Tashiro}\altaffilmark{5},
  Yasushi \textsc{Fukazawa}\altaffilmark{6},
  Tuneyoshi \textsc{Kamae}\altaffilmark{9},
  Greg M. \textsc{Madejski} \altaffilmark{9},\\
  Kazuhiro \textsc{Nakazawa}\altaffilmark{3},
  Kazutaka \textsc{Yamaoka}\altaffilmark{7},
  Yukikatsu \textsc{Terada}\altaffilmark{2},
  Daisuke \textsc{Yonetoku}\altaffilmark{4},\\
  Shin \textsc{Watanabe}\altaffilmark{3},
  Toru \textsc{Tamagawa}\altaffilmark{2},
  Tsunefumi \textsc{Mizuno}\altaffilmark{6},
  Aya \textsc{Kubota}\altaffilmark{2},\\
  Naoki \textsc{Isobe}\altaffilmark{2},
  Isao \textsc{Takahashi}\altaffilmark{1},
  Goro \textsc{Sato}\altaffilmark{3},
  Hiromitsu \textsc{Takahashi}\altaffilmark{6},\\
  Soojing \textsc{Hong}\altaffilmark{8},
  Madoka \textsc{Kawaharada}\altaffilmark{1},
  Naomi \textsc{Kawano}\altaffilmark{6},
  Takefumi \textsc{Mitani}\altaffilmark{3},\\
  Mio \textsc{Murashima}\altaffilmark{1},
  Masaya \textsc{Suzuki}\altaffilmark{5},
  Keiichi \textsc{Abe}\altaffilmark{5},
  Ryouhei \textsc{Miyawaki}\altaffilmark{1},\\
  Masanori \textsc{Ohno}\altaffilmark{6},
  Takaaki \textsc{Tanaka}\altaffilmark{3,1},
  Takayuki \textsc{Yanagida}\altaffilmark{1},
  Takeshi \textsc{Itoh}\altaffilmark{1},\\
  Kousuke \textsc{Ohnuki}\altaffilmark{3,1},
  Ken-ichi \textsc{Tamura}\altaffilmark{3,1},
  Yasuhiko \textsc{Endo}\altaffilmark{5},
  Shinya \textsc{Hirakuri}\altaffilmark{1},\\
  Tatsuro \textsc{Hiruta}\altaffilmark{3},
  Takao \textsc{Kitaguchi}\altaffilmark{1},
  Tetsuichi \textsc{Kishishita}\altaffilmark{3,1},
  Satoshi \textsc{Sugita}\altaffilmark{7},\\
  Takuya \textsc{Takahashi}\altaffilmark{6},
  Shin'ichiro \textsc{Takeda}\altaffilmark{3,1},
  Teruaki \textsc{Enoto}\altaffilmark{1},
  Ayumi \textsc{Hirasawa}\altaffilmark{6},\\
  Jun'ichiro \textsc{Katsuta}\altaffilmark{3,1},
  Satoshi \textsc{Matsumura}\altaffilmark{5},
  Kaori \textsc{Onda}\altaffilmark{5},
  Mitsuhiro \textsc{Sato}\altaffilmark{1},\\
  Masayoshi \textsc{Ushio}\altaffilmark{3,1},
  Shin-nosuke \textsc{Ishikawa}\altaffilmark{3,1},
  Koichi \textsc{Murase}\altaffilmark{5},
  Hirokazu \textsc{Odaka}\altaffilmark{3,1},\\
  Masanobu \textsc{Suzuki}\altaffilmark{5},
  Yuichi \textsc{Yaji}\altaffilmark{5},
  Shinya \textsc{Yamada}\altaffilmark{1},
  Tomonori \textsc{Yamasaki}\altaffilmark{6},\\
  Takayuki \textsc{Yuasa}\altaffilmark{1},
  and the HXD team}
\altaffiltext{1}{Department of Physics, University of Tokyo, 7-3-1 Hongo, Bunkyo-ku, Tokyo 113-0033}
\email{kokubun@phys.s.u-tokyo.ac.jp}
\altaffiltext{2}{Cosmic Radiation Laboratory, The Institute of Physical and
Chemical Research (RIKEN),\\
2-1 Hirosawa, Wako, Saitama 351-0198}
\altaffiltext{3}{Department of High Energy Astrophysics, Institute of 
Space and Astronomical Science (ISAS),\\
Japan Aerospace Exploration Agency (JAXA),
3-1-1 Yoshinodai, Sagamihara, Kanagawa 229-8510}
\altaffiltext{4}{Department of Physics, Kanazawa University,
Kakuma, Kanazawa, Ishikawa 920-1192}
\altaffiltext{5}{Department of Physics, Saitama University, \\
Shimo-Okubo, Sakura-ku, Saitama-shi, Saitama 338-8570}
\altaffiltext{6}{Department of Physics, Hiroshima University,\\
1-3-1 Kagamiyama, Higashi-Hiroshima, Hiroshima 739-8526}
\altaffiltext{7}{Department of Physics and Mathematics, 
Aoyama-gakuin University, \\
5-10-1 Fuchinobe, Sagamihara, Kanagawa 229-8558}
\altaffiltext{8}{College of Science and Technology, Nihon University,\\
7-24-1 Narashinodai, Funabashi-shi, Chiba 274-8501}
\altaffiltext{9}{Stanford Linear Accelerator Center, 2575 Sand Hill Road, 
Menlo Park, CA 94025, USA}


\maketitle

\begin{abstract}
The in-orbit performance and calibration of the Hard X-ray Detector (HXD)
on board the X-ray astronomy satellite Suzaku are described.
Its basic performances, 
including a wide energy bandpass of 10--600 keV,
energy resolutions of $\sim$4 keV (FWHM) at 40 keV 
and $\sim$11\% at 511 keV, 
and a high background rejection efficiency, 
have been confirmed by extensive in-orbit calibrations. 
The long-term gains of PIN-Si diodes have been stable
within 1\% for half a year, 
and those of scintillators have decreased by 5-20\%.
The residual non-X-ray background of the HXD is 
the lowest among past non-imaging hard X-ray instruments
in energy ranges of 15--70 and 150--500 keV.
We provide accurate calibrations of energy responses, angular responses, 
timing accuracy of the HXD, and relative normalizations to
the X-ray CCD cameras using multiple observations of the Crab Nebula.
\end{abstract}

%
%
\section{Introduction}
\label{section:1}

The fifth Japanese X-ray satellite, Suzaku, was launched on 2005
July 10 into a low earth orbit of $\sim$570 km altitude 
and 32$^{\circ}$
inclination \citep{Mitsuda2006}. The satellite carries four X-ray 
CCD cameras (X-ray Imaging Spectrometer - XIS; \cite{Koyama2006}),
which are placed at the focal points of the four X-ray 
telescopes (XRT; \cite{Serlemitsos2006}) and
covers the soft energy range of 0.2--12 keV. 
The satellite also carries a non-imaging
hard X-ray instruments, the Hard X-ray Detector (HXD), which is the
subject of the present paper. 
The detailed design of the experiment
and basic performances in the pre-launch calibration are 
described by \authorcite{Takahashi2006} (\yearcite{Takahashi2006}; 
hereafter Paper I), 
followed by brief descriptions of the initial in-orbit performance
by \authorcite{Fukazawa2006} (\yearcite{Fukazawa2006})
and \authorcite{Kitaguchi2006} (\yearcite{Kitaguchi2006}).

The HXD consists of 
three parts contained in separate chassis: the sensor (hereafter HXD-S), 
the analog electronics (HXD-AE), and the digital electronics (HXD-DE).
The HXD achieves an extremely low detector background through a highly 
ingenious structure of HXD-S, a compound-eye configuration of 
4$\times$4 well-type phoswich units (``Well units'') surrounded by 20 
thick active shields (``Anti units''). 
In addition to signals from all of 36 units, those from 64 PIN-Si 
diodes inside the well-type phoswiches are also fed into the parallel 
readout system in HXD-AE, and the hard-wired anti-coincidence system 
drastically reduces the detector background by use of the hit-pattern 
signal from active shields. Further intelligent event screenings are
realized by the onboard software in HXD-DE (Paper I). 

Extensive in-orbit calibrations for all the hundred signals are crucial,
to confirm that the instrument survived the launch, to optimize
the hardware/software settings and the daily operation scheme of 
the HXD, and to verify the detector performance in orbit.
We summarize the in-orbit operations in section 2.
The in-orbit performances of PIN-Si diodes (hereafter PIN) and 
the gadolinium silicate scintillators (Gd$_2$SiO$_5$:Ce, hereafter GSO)
are described in section 3 and 4, respectively. 
The spectral and temporal properties
of the residual background are explained in section 5.
In section 6, we address other 
miscellaneous calibration issues including the angular response, dead time
estimation, and timing accuracy.

%
%
\section{Initial Operation of the HXD}
\label{section:2}

On 2006 July 22, about two weeks after the launch,
the run-up operation of the electronic system of HXD started.
It took a few days to turn on the low-voltage part of the experiment,
and upload initial settings of the onboard hardware and software.
After that, an extended period of high-voltage turn-on
followed over a week, in which the parameter tuning of the electronics
was also performed. 
These operations are summarized in table~\ref{tbl:runuplog}.

\begin{longtable}{ll}
  \caption{Run-up operation procedures of the HXD.}
  \label{tbl:runuplog}
  \hline\hline
      Date & Operation \\
\endfirsthead
  \hline\hline
      Date & Operation \\
\endhead
  \hline
\endfoot
\endlastfoot
      \hline
      July 11--13  ....& HXD-S temperature 20 $\rightarrow$ -15 $^{\circ}$C\\
      July 22  ........... & HXD on, HXD-DE on, CPU run, Observation mode \\
                      & HXD-S temperature -15 $\rightarrow$ -20 $^{\circ}$C\\
      July 25  ........... & WPU0-3 on, TPU0-3 on, HXD-AE initial parameter load \\
      July 27  ........... & HV-W0-3, HV-T0-3, HV-P0-3 on \\
      July 27--Aug. 4 & HV operation, AE/DE parameter tuning \\
      Aug. 8--15  ........ & HV reduction \\
      Aug. 15--18  ...... & HV operation, AE/DE parameter tuning \\
      Aug. 19  ........... & First-light (CenA)\\
      \hline
\end{longtable}

\subsection{Temperature Control of HXD}
\label{subsection:2-1}

The heat generated from the electrical power consumption 
in HXD-S is transported through two sets of heat-pipes, 
which are thermally connected to the ``cold plate'' beneath HXD-S,
and then released from two radiators on the spacecraft side panels 
number 6 and 8 \citep{Mitsuda2006}. The cooling is compensated by two pairs of heaters 
which are also attached to the cold plate. Thus, the temperature of 
HXD-S is designed to be controlled in the orbit within $-20\pm5$ C
(Paper I), which is the optimum for low thermal noise in PIN 
and high light yields in scintillators.

Since a large temperature gradient within HXD-S 
would give excess thermal strain to the scintillators, 
the HXD-S temperature should be changed gradually, 
by no more than a limit of 5 degree per hour. 
On the next day of the launch, 
solar array panels were opened and temperatures
of the instruments inside the spacecraft started decreasing rapidly.
The temperature of the cold plate of HXD-S was thus controlled 
to drop from 20 to $-15$ $^{\circ}$C with a step of 5 or 2.5 degree, 
during contact passes of the succeeding three days.
Then, it was further lowered to the nominal operation temperature,
$-20$ C, after the turn-on of HXD-DE.

\begin{figure}
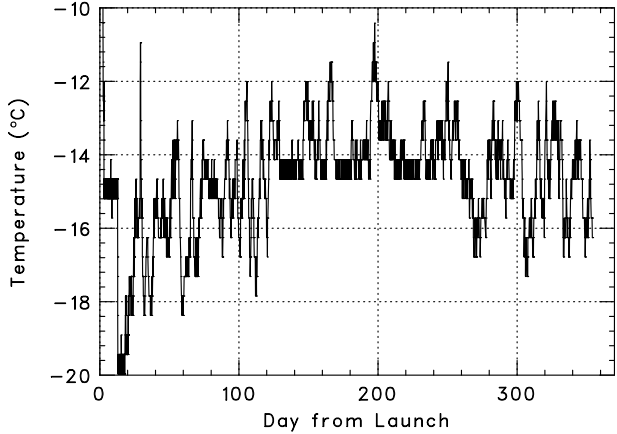

\begin{center}
\FigureFile(0.48\textwidth,0.48\textwidth){figure1.eps}
\end{center}
\caption{The temperature history of HXD-S measured at the cold plate
for about a year from the launch. 
The control temperature was kept at $-14.5$ $^{\circ}$C from 129 to 255 days 
after the launch.}
\label{fig:temp_history}
\end{figure}

In the very early phase of this temperature control operation, 
it was found that one of the two heat-pipes, 
connecting to the radiator on side panel 6,
was not functional, implying that the available heat transport 
capability became half the designed value.
As a result, the actual temperature of HXD-S remained
around $-16$ C, even though the control temperature was 
set to $-20$ C. Furthermore, the temperature could not be 
``controlled'' to sufficiently low values, and hence it
fluctuated by attitude changes of the satellite, which 
affected solar heat in-flows to the spacecraft.
To measure long-term gain variations of photo-multiplier
tubes (PMTs) free of temperature-dependent gain changes,
the control temperature was changed 
to $-14.5$ $^{\circ}$C on 2005 November 16  until 2006 March 22.
Although it was feared that the higher HXD-S temperature would 
enhance thermal noise of the PIN diodes, 
the effect was confirmed to appear only at lower energy
range than 10--12 keV (\S\ref{subsection:3-2}), 
as long as the temperature is below $-11$ C.
The resultant temperature light curve of HXD-S,
measured at the cold plates, 
is shown in figure~\ref{fig:temp_history}.

\subsection{High-Voltage Operations}
\label{subsection:2-2}

HXD-S uses four high-voltage supply units for PIN diodes (PIN-HV)
which can provide up to 600 V, and eight units for PMTs
(PMT-HV) up to 1250 V. One PMT-HV drives four PMTs
that share the same electronics module (WPU or TPU; Paper I),
while one PIN-HV supplies bias voltages for 16 diodes.
When increasing a PIN-HV output, a step increment of less than
100 V is used not to destroy FETs inside the charge sensitive 
pre-amplifiers.

The high-voltage run-up operation
started after the 17 day waiting period to let the spacecraft
to fully outgas, and to prevent the high voltages from discharging.
The PIN-HVs were first operated at low voltages (50 V) for short
intervals ($\sim$10 min) when the satellite was on contact
from the tracking station. Even with this low bias voltage,
large pulse-height events, caused by cosmic-ray particles 
penetrating the detector, 
were observed at a rate of $\sim$ 10 ct s$^{-1}$ 
(summed over the 4 PIN diodes in the same phoswich unit).
Using such particle events, all the 64 diodes were confirmed to 
have survived the severe launch vibration.
Then, over about a week period, 
the operation voltage and its duty cycle
were gradually increased to the nominal value of $\sim$500 V and 100\%.

\begin{table}
\caption{Nominal high voltages.}
\label{tbl:hvsetting}
\begin{center}
\begin{tabular}{lllll}
  \hline\hline
      HV unit \# & 0 & 1 & 2 & 3 \\
      \hline
      PIN & 489 & 489 & 490 & 490 \\
      PMT (Well) & 850 & 872 & 875 & 902 \\
      PMT (Anti) & 816 & 860 & 878 & 874 \\
      \hline
\end{tabular}
\end{center}
\end{table}

After confirmation of the normal functioning of the PIN diodes and PIN-HVs, 
the output voltage from PMT-HVs were also increased to 500 V.
At this stage, 
all PMT units have been confirmed to be properly functioning,
by use of the cosmic-ray events. In addition, the anti-coincidence
particle reduction with the hit-pattern signals was also confirmed
to be working as designed.
The operation voltages and the duty cycle were gradually increased
in the same manner as the PIN-HV, up to individual nominal voltages
which were determined from the pre-launch gain measurements.
Table~\ref{tbl:hvsetting} summarizes the achieved final 
operation voltages for all high-voltage units, which have remained 
unchanged throughout the performance verification phase.

The outputs from PMT-HVs are all reduced to zero by programmed
commands during the South Atlantic Anomaly (SAA) passages, in which 
huge number of charged particles hit the detector. 
This manual operation is backed up with an automatic control by 
radiation belt monitor (RBM) function, based on the counting rates
of four corner shield units (Paper I; \cite{Yamaoka2006}). 
The RBM flag was sometimes triggered by 
intense solar flares in the early phase, 
and it was confirmed that the reduction sequence works properly.
While a nominal counting rate of one corner unit is $\sim$1000 ct s$^{-1}$,
it is expected to reach a few 10 kct s$^{-1}$
when the PMT-HVs are accidentally not reduced during 
the SAA passage, 
and the actual counting rate in solar-side two units recorded 
more than 25 kct s$^{-1}$ in an X2.0 class solar flare.
Therefore, the threshold rates of the RBM function were finalized 
as 100 kct s$^{-1}$ for solar-side units (T00, T10) 
and 10 kct s$^{-1}$ for others (T20, T30).

\subsection{Electronics Setup}
\label{subsection:2-3}

HXD-AE has various adjustable parameters, which can be changed by
commands for individual detector units (Paper I). On July 25, HXD-AE 
was loaded tentatively with a nominal parameter table, which was 
determined based on the ground calibration. After the high-voltage 
operation was mostly completed, each parameter was re-optimized according 
to the in-orbit data. For both PIN and GSO, the hardware event selection
with lower and upper discriminators (LD and UD) and pulse-shape
discrimination (PSD) was set to be as loose as possible, 
provided that the data transfer rate from HXD-AE to HXD-DE
stays within the hardware limit of 128 kbps ($\sim$1000 events s$^{-1}$) 
per WPU (Paper I).
The achieved final parameters are summarized in table~\ref{tbl:aesetting}.

In the case of PIN, pulse heights from all the 64 diodes are 
adjustable with a common gain for every four gain-amplifiers, while 
trigger signals are produced at comparators which have also a common 
threshold voltage in each WPU. Then, the triggers produce corresponding
event records with a sampling resolution of 8-bit (256 bins).
The gains and LD levels have been kept almost the same 
as the nominal ones in all the PIN diodes, because their performance did not
change significantly after the launch.
The final settings allow a dynamic range of 8$\sim$90 keV 
which completely satisfies the design goals, 
and the digitization of 256 channel pulse-height
spectrum, $\sim$0.4 keV per channel, are fine enough for 
the typical energy resolution of PIN ($\sim$4 keV).

The counting rates of the LD and UD from every four PIN diodes 
(PIN-LD and PIN-UD) in a same Well unit, recorded by scalers in HXD-AE, 
are edited every telemetry period (nominally 2 or 4 s) into the house 
keeping (HK) data, and utilized to monitor the raw trigger rates 
before the onboard event reduction. 
While the PIN-LD rate from most of 16 units stay within 
10$\sim$50 ct s$^{-1}$ in orbit, 
some units sometimes exhibit exceptionally high counting
rates of $>$100 ct s$^{-1}$, especially during the daytime of the satellite. 
This is thought to be caused by the electrical interference from a large 
surplus current of the solar paddle, 
which are dissipated at the shunt resistors on the side-panel.

Although the PIN-Si diodes employed in the HXD have 
an unprecedented thickness of 2 mm (Paper I),
they have negligible ($\le$1\%) 
cross-sections for hard X-ray photons with energies
higher than the UD level ($\sim$90 keV);
therefore,  the counting rate of PIN-UD can 
be regarded as the number of cosmic-ray charged particles penetrating the device. 
A typical in-orbit rate of PIN-UD is $\sim$10 ct s$^{-1}$ per Well unit, 
corresponding to $\sim$1 particle s$^{-1}$ cm$^{-2}$.
As mentioned later (\S\ref{subsection:5-1}),
this method can be also applied to estimate the particle flux
during the SAA, since the high-voltages for PIN are always kept on.

In contrast to the case of PIN, various in-orbit fine tunings were 
necessary for GSO, mainly because PMT gains can generally, and did
actually, change by up to $\sim$10\% due to the launch vibration.
Relative gains of the 16 Well units were first 
adjusted by trimming gain-amplifiers for the ``slow'' shaping signals
in HXD-AE, using the intrinsic natural 
radio-active isotope (\atom{Gd}{}{152}; Paper I).
The ``fast'' 
gains were also trimmed for the GSO events to have the same pulse heights
as their ``slow'' pulse heights. As a result, the slope of GSO branch 
on the 2-dimensional fast-slow diagram became close to diagonal,
and hence the hardware PSD cut can utilize the same conditions as those
optimized on ground.

The LD levels of the anode trigger (anode LD) were 
set for individual units at around 30 keV, while the UD levels 
were set at $\sim$900 keV, shared by four units. 
Another lower threshold, called ``slow LD'' (SLD), 
which is applied to the slow shaped signal to generate
hit-pattern flags, was kept at the pre-launch value in all units.
The SLD level corresponds to an energy deposit of $\sim$20 keV if
it occurs in GSO, $\sim$50 keV if in the bottom block of the BGO 
shield, and $\sim$100 keV if in the Well-shaped BGO top part.
With these settings, typical counting 
rates of LD, UD, and SLD from one unit are
700$\sim$1000 ct s$^{-1}$, 50$\sim$100 ct s$^{-1}$, 
and 1000$\sim$1500 ct s$^{-1}$, respectively.
Although each anode trigger initiates a data acquisition sequence,
most of them are immediately rejected before the analog-to-digital
conversion stage when the hard-wired PSD function is enabled, 
and the rate of events acquired as digitized data packets is 
successfully reduced to be less than $\sim$100 ct s$^{-1}$.

Widths of the hit-pattern signals from both Well and Anti units
are also adjustable from 4.2 to 5.6 $\mu$s by commands (Paper I). 
A longer width usually yields a higher reduction efficiency of 
the anti-coincidence, at the sacrifice of an increase of the 
accidental coincidence. Two different widths of 4.6 and 5.6 $\mu$s
were tested during an observation of a blank sky field
to investigate an optimum in the orbit.
Since the longer one yielded significantly ($\sim$20\%) lower 
background of PIN with only a small increase ($\sim$1\%)
of the accidental coincidence, the latter was employed.
Widths of trigger generation vetoing,
which suppresses false triggers after large signals above the UD level, 
were fixed to be the same as that optimized on ground (40 $\mu$s).

\begin{longtable}{llll}
\caption{Nominal setup of HXD-AE.}
\label{tbl:aesetting}
\hline\hline
 & Setting & Energy\footnotemark[$*$] & Common Units\footnotemark[$\dagger$]\\
\endfirsthead 
\hline\hline
 & Setting & Energy\footnotemark[$*$] & Common Units\footnotemark[$\dagger$]\\
\endhead
\hline
\endfoot
\endlastfoot
\hline
PIN & & &\\
\hline
Gain.....................................& $\times$5.0  & & 4 PIN \\
Analog LD...........................& 68--85 mV  & $\sim$6--8 keV & 16 PIN \\
\hline
PMT & & &\\
\hline
Slow Gain............................& $\times$2.0--2.4 & & individual\\
Fast Gain.............................& $\times$2.2--2.5 & & individual\\
PSD Level...........................& 300--400 mV& $\sim$30--40 keV  & 4 Well \\
Anode LD Level..................& 30--40 mV & $\sim$25--30 keV & individual\\
SLD Level............................& 125--133 mV  & $\sim$15--20 keV  & individual\\
UD Level..............................& 2.06 V       & $\sim$800--900 keV& 4 Well \\
UD Veto Width...................& 40 $\mu$s &     & 4 Well \\
Hit-pattern Width...............& 5.6 $\mu$s &  & 4 Well \\
\hline
\multicolumn{4}{l}{\hbox to 0pt{\parbox{180mm}{\footnotesize
\par\noindent
\footnotemark[$*$] Rough conversion into the energy.
\par\noindent
\footnotemark[$\dagger$] Number of PIN diodes or Well units, 
which are commonly applied with the same parameter.
}\hss}}
\end{longtable}

\subsection{Onboard Software Setup}
\label{subsection:2-4}

Even after the hard-wired PSD cut, the total event rate summed over 
the four WPU modules typically reaches a few kct s$^{-1}$, 
significantly higher than the nominal telemetry limit ($\sim$300 ct s$^{-1}$),
mainly due to the electrical interferences and insufficient 
temperature control after the launch.
However, the onboard software in HXD-DE have been designed to 
be flexible with its various event selection functions,
and hence further event reductions can be achieved.
The onboard software can judge the events based on the
PIN or GSO pulse heights, as well as subsidiary
information such as the trigger pattern, hit-pattern, and quality flags 
(Paper I) contained in each event data.

The PIN event data, sent from HXD-AE to HXD-DE, contain
cosmic-ray produced saturated events at a rate 
of $\sim$10 ct s$^{-1}$ per one well unit; 
these can be easily removed by use of the 
PIN-UD flag. In addition, the electric interference from the satellite 
power line was observed in some units, at a rate up to $\sim$100 ct s$^{-1}$.
Since the interference-produced events appear as common-mode noise
among neighboring PIN diodes, they can be eliminated in HXD-DE
based on multiple triggers among the four PIN diodes in the 
same Well unit; that is, 
PIN events with single trigger are out to the telemetry.
Finally, tighter lower threshold levels are digitally applied to 
individual PIN signals,
to remove thermal noise events with low pulse heights. 
This ``digital LD'', ranging 16--32 ADC channels, 
can be individually applied by commands to the 64 PIN pulse-heights, 
whereas the analog LD in HXD-AE is common among the sixteen
PIN diodes in the same WPU module.
As a result, the average of summed event rate from 64 PINs is 
ordinarily reduced down to 6--10 ct s$^{-1}$, 
although there still remains
a rapid increase up to 40 ct s$^{-1}$ during the daytime of the satellite.

\begin{figure*}
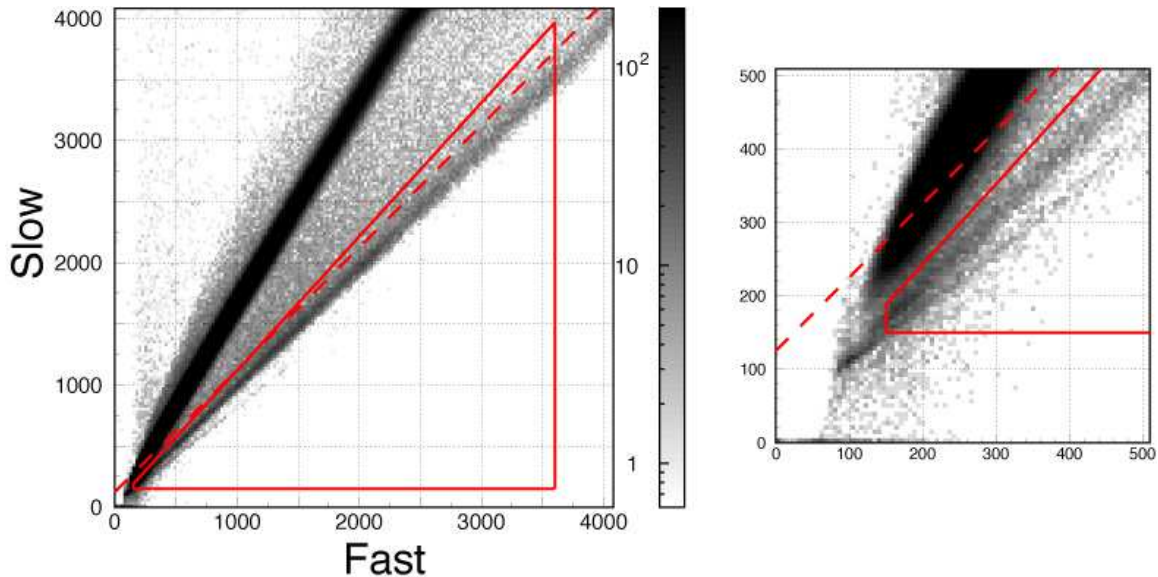

\begin{center}
\FigureFile(0.90\textwidth,0.48\textwidth){figure2.eps}
\end{center}
\caption{The digital PSD selection criteria (solid trapezoid), 
shown on a two-dimensional histogram of fast and slow shaped pulse heights, 
obtained while the hard-wired PSD function is disabled.
The dashed line denotes the hard-wired PSD cut. 
}
\label{fig:gso_psdcut}
\end{figure*}

To suppress the GSO event rate, a tighter setting in the PSD cut 
condition is inevitable,
because the digitized ``GSO'' data actually contain a large 
number of BGO events, particularly at the lower energy end.
As shown in figure~\ref{fig:gso_psdcut}, 
individual trapezoids are defined on the fast-slow histograms,
to discard the residual BGO branch.
These boundaries, called ``digital PSD'' hereafter, 
at the same time eliminate events with too low or
too high pulse heights, 
corresponding to a digital lower
and upper discriminator at $\sim$30 keV and $\sim$700 keV, respectively.
The summed GSO event rate from the 16 Well units is thus 
reduced to 70--150 ct s$^{-1}$,
which varies according to the satellite position in the orbit
(\S\ref{subsection:5-2}).
In table \ref{tbl:evreduction}, 
the counting rate reduction of PIN and GSO
by each stage of cut in HXD-AE, HXD-DE, and analysis software
(\S\ref{subsection:3-3}, \ref{subsection:4-4}, and \ref{subsection:4-5})
is summarized.

\begin{longtable}{lcc}
\caption{Counting rate of PIN and GSO at each stage of cut.}
\label{tbl:evreduction}
\hline\hline
 & PIN (ct s$^{-1}$ unit$^{-1}$)\footnotemark[$*$] 
 & GSO (ct s$^{-1}$ unit$^{-1}$)\footnotemark[$*$] \\
\endfirsthead 
\hline\hline
 & PIN (ct s$^{-1}$ unit$^{-1}$) & GSO (ct s$^{-1}$ unit$^{-1}$)\\
\endhead
\hline
\endfoot
\endlastfoot
\hline
HXD-AE & &\\
\hline
Initial analog LD rate........................& 10--100  & 700--1000\\
After analog PSD cut........................& -------- &  $<$100\\
\hline
HXD-DE & &\\
\hline
After PIN UD cut..............................& 1--100   & --------\\
After single PIN trigger cut...............& 1--10    & --------\\
After digital LD.................................& 0.4--2.5 & --------\\
After digital PSD...............................& -------- &  5--10\\
\hline
Analysis Software\footnotemark[$\dagger$] & &\\
\hline
After anti-coincidence applied............& 0.025--0.075 &  1.5--2.5\\
\hline
\multicolumn{3}{l}{\hbox to 0pt{\parbox{180mm}{\footnotesize
\par\noindent
\footnotemark[$*$] Counting rate per a Well unit.
\par\noindent
\footnotemark[$\dagger$] Described in 
\S\ref{subsection:3-3}, \ref{subsection:4-4}, and \ref{subsection:4-5}.
}\hss}}
\end{longtable}

\subsection{Operation History}
\label{subsection:2-5}

After the initial run-up operation, the parameters of HXD-AE and
the event selection conditions in HXD-DE have been basically kept
unchanged throughout the performance verification phase, 
except for some minor adjustments summarized in table~\ref{tbl:opelog}.
In addition, the nominal observation of HXD was a few times 
interrupted for a purpose of the memory dump operation of HXD-DE,
to investigate unexpected status errors in-orbit.

\begin{longtable}{ll}
\caption{Major operations of the HXD during the first year.}
\label{tbl:opelog}
\hline\hline
Date & Operation \\
\endfirsthead 
\hline\hline
Date & Operation \\
\endhead
\hline
\endfoot
\endlastfoot
\hline
Sep.17.......... & HXD-AE RBM level raised : 10$\rightarrow$100 kHz \\
Oct.22.......... & HXD-DE memory dump \\
Nov.16.......... & HXD-S control temperature raised : -20$\rightarrow$-14.5 $^{\circ}$C \\
Dec.30--Jan.4    & PMT-HV reduced to 0 V \\
Mar.20.......... & WAM time resolution changed : 1.0$\rightarrow$0.5 sec \\
Mar.22.......... & HXD-S control temperature lowered : -14.5$\rightarrow$-20 $^{\circ}$C \\
Mar.23.......... & GSO anode LD lowered : 40$\rightarrow$30 mV \\
 & Digital PSD range changed : 150--3600$\rightarrow$120--3000 ch \\
                 & Telemetry rate at Bit-M changed : 33$\rightarrow$44 kbps\\
Mar.27.......... 
 & Digital PSD range changed : 120--3000$\rightarrow$100--3000 ch \\
Apr.15.......... & Telemetry rate at Bit-L changed :10$\rightarrow$15 kbps\\
May.13.......... & GSO anode LD raised : 30$\rightarrow$40 mV\\
May.17.......... & HXD-DE memory dump\\
\hline
\end{longtable}

An ordinary daily operation of the HXD includes a fixed sequence of
health check, resetting counters, and outputting diagnostic information.
In the nominal observation mode, 
the hardware and software settings are optimized to reduce the
detector background as much as possible.
However, background events are still useful for diagnostic purposes.
Therefore, by utilizing 
earth occulted periods in the orbit, the PSD selection in both
of HXD-AE and HXD-DE are disabled for 10 minutes every day, 
to monitor the BGO events from individual Well units. 
At the same time, the digital thresholds of PIN events 
are also disabled to obtain noise spectra from individual
PIN diodes. Therefore, a much higher event rate than the maximum
transfer rate saturates the telemetry during this period, and
prevent the instrument from performing any scientific observation.

Each WPU module has time counters to record latched values on
each triggered event. These counters are simultaneously reset 
at the beginning of each observation by programmed commands. 
Counters for LD and UD are reset at the end of every SAA passage, 
to restart  anode LD counters 
which sometimes ``freeze'' due to a bug of digital logic in WPU.

\subsection{Data Processing}
\label{subsection:2-6}

Both scientific and house keeping data are immediately recorded 
in an onboard data recorder, and later transmitted to the ground 
tracking stations as raw telemetry packets. These data are promptly
transferred to storages in the data center of ISAS/JAXA,
and converted into the Flexible Image Transport System 
(FITS; \cite{Wells1981}) data format.
Standard pipeline processings are then applied, which consist
of following data handlings with the relevant software 
(FTOOLS; \cite{Blackburn1995}).
Throughout this paper, we have used pipeline products, 
processed with a set of softwares and calibration files 
tagged as a version 1.2
\footnote{\tt http://www.astro.isas.jaxa.jp/suzaku/process/history/v1223.html}.

First, the absolute timing is assigned by {\tt hxdtime}, relying
on time record of 19 bits length and 61 $\mu$s precision 
(Paper I) in individual events, and on coarse timing information 
contained in a header block of event data packets. An absolute
time for each event is obtained by homologizing the time record
to the original clock in the satellite digital processor, and
then synchronizing it to the standard oscillator on the ground.
Transmission delays and temperature drift of the onboard clock
are also taken into account in this process.
In the next step, pulse-heights of each event are transformed by {\tt hxdpi},
from analog-to-digital converter (ADC) channels into pulse-height 
invariants (PI), 
which is linearly proportional to the physical energy of incident photons.
Energy range and channel numbers of PI are commonly defined as
0.375--96.375 keV with 256 channels for PIN events,
and as 1--1025 keV with 512 channels for GSO events.
In case of PIN events, fixed conversion factors are applied 
on the 64 pulse-heights to get the proper energy scale, 
using the in-orbit calibration (\S\ref{subsection:3-1}) and
non-linearity corrections in HXD-AE.
On the other hand, 
in the case of GSO events, due to temperature-dependent PMT gain changes,
time-dependent conversion factors are applied by in-orbit
calibration lines (\S\ref{subsection:4-1}) to get the proper energy scale.
Finally, events are qualified by {\tt hxdgrade}, based on the 
trigger patterns, hit-pattern flags, and various quality information
recorded by HXD-DE (Paper I). 
The selection criteria optimized for the PSD selection 
(\S\ref{subsection:4-4}) and anti-coincidence 
(\S\ref{subsection:3-3}, \S\ref{subsection:4-5}) are also applied, 
and events which passed all of these cuts
are tagged as ``clean events''.

%
%
\section{In-orbit Performance of PIN}
\label{section:3}

In-orbit calibrations of the PIN diodes have been carried out in 
three steps. 
First, for each of the 64 PINs, 
the absolute energy scale is established,
the energy resolution (\S\ref{subsection:3-1}) is evaluated,  
and the lower energy threshold (\S\ref{subsection:3-2}) is optimized.
Second, event selection criteria is optimized so as to minimize 
the residual non X-ray background (\S\ref{subsection:3-3}).
Finally, the response matrices of individual PINs are constructed,
based on quantum efficiencies and effective areas
(\S\ref{subsection:3-4}).

\subsection{Energy Scale}
\label{subsection:3-1}

\begin{figure}
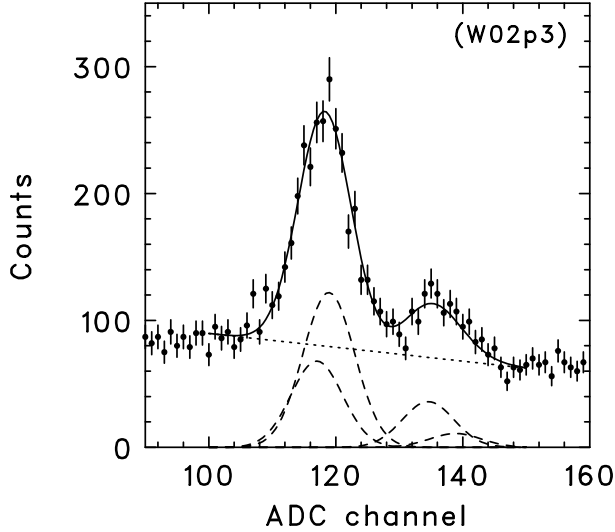

\begin{center}
\FigureFile(0.48\textwidth,0.48\textwidth){figure3.eps}
\end{center}
\caption{An energy spectrum of a single PIN diode, in which the coincident 
events of PIN and GSO are accumulated over a half year.
Four gaussians indicated by dashed lines correspond to 
K$_{\alpha1}$ (43.0 keV), K$_{\alpha2}$ (42.3 keV), 
K$_{\beta1+\beta3}$ (48.6 keV), and K$_{\beta2}$ (50.0 keV) 
fluorescence lines of gadolinium, 
while the linear component shown by the dotted line denotes 
a background continuum.}
\label{fig:pin_gdk}
\end{figure}

Before the launch, the energy scales of the 64 PINs were precisely 
measured using the standard $\gamma$-ray sources, within 
$\sim$1\% accuracy (Paper I). 
These energy scales, or gains, are not expected to change significantly
after the launch, since neither charge collection efficiency of the
PIN diodes nor capacitance of the charge sensitive amplifiers is 
sensitive to the environmental changes. Nevertheless, the energy scale
is so important that it should be accurately reconfirmed using the
actual data.
Instead of the calibration isotopes, fluorescent X-rays from gadolinium
(Gd-K$_\alpha$ 42.7 keV) and bismuth (Bi-K$_\alpha$ 76.2 keV) in GSO 
and BGO scintillators, respectively, 
can be used for in-orbit energy scale calibrations.
These fluorescent events are hardly detected in the "clean events",
because they are eliminated by the anti-coincidence with the scintillator
signals, caused by simultaneous energy deposits.
Consequently, 
as shown in figure~\ref{fig:pin_gdk},
they can be extracted with a high signal-to-noise
ratio by selecting only coincidence events between PIN diodes and
scintillators of the same Well unit.

\begin{figure}
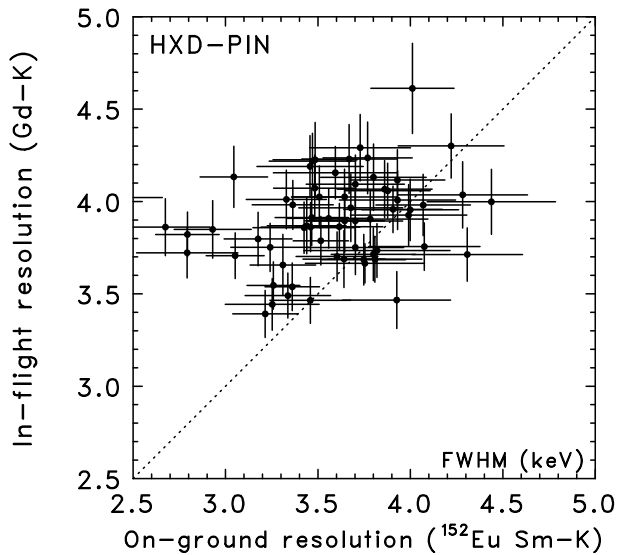

\begin{center}
\FigureFile(0.48\textwidth,0.48\textwidth){figure4.eps}
\end{center}
\caption{Comparison between the energy resolutions of the PIN diodes
measured on-ground and in-orbit, by use of \atom{Eu}{}{152} isotope 
(Sm-K$_{\alpha1}$ : 40.1 keV) and the Gd-K$_{\alpha1}$ (43.0 keV) line, 
respectively.}
\label{fig:pin_eneres}
\end{figure}

As shown in figure~\ref{fig:pin_gdk}, 
by fitting the pulse height spectra with four Gaussians,
which represent K$_{\alpha1}$, K$_{\alpha2}$, K$_{\beta1+\beta3}$, 
and K$_{\beta2}$ transition lines and a background continuum,
the peak channels of K$_{\alpha1}$
line were obtained for individual PINs, and the energy resolution of 
that peak was measured at the same time. 
In this fitting, the four Gaussian centroid energies were constrained
to obey theoretical line-energy ratios, and the Gaussian widths were
tied together but left a free parameter.
In figure~\ref{fig:pin_eneres},
thus obtained energy resolutions of the 64 PINs are plotted against 
those measured in the pre-launch calibration. 
The typical in-orbit energy resolution for the Gd-K$_{\alpha}$ 
line is obtained as $\sim$4 keV in FWHM, which is roughly consistent
with those measured before the launch.
A slight increase of $\sim$0.3 keV from pre-launch resolutions
are probably due to a difference in the electrical noise conditions.

\begin{figure}
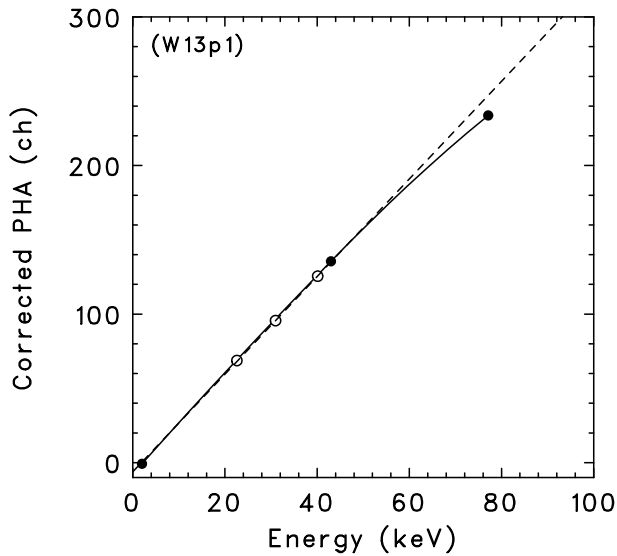

\begin{center}
\FigureFile(0.48\textwidth,0.48\textwidth){figure5.eps}
\end{center}
\caption{
Energy vs. pulse-height linearity of a representative PIN diode.
Open and filled circles represent the on-ground and in-orbit
calibration points, respectively.
The solid line indicates the spline curve obtained with 
the in-orbit calibration points, while the dashed line
shows the linear energy scale determined with the on-ground 
measurements. 
The pulse height is shown in a unit of pulse-height amplitude
(PHA), which is equivalent to the ADC channel after the
correction of nonlinearity in HXD-AE.}
\label{fig:pin_escale}
\end{figure}

In addition to the two fluorescent lines, another anchor point is 
needed at low energies to accurately fix the energy scales.
For this purpose, ``{\it pedestal channel}'', which is defined as 
the peak channel of noise spectrum obtained by the random triggers 
from scintillators, is used. 
Although the energy deposits to the relevant PIN diode are 
considered in this case essentially zero, the channel becomes
non-zero, because the peak-hold circuit before ADC latches the
noise peak during each trigger gate of a $\sim$10 $\mu$s width.
Therefore, the {\it pedestal channel} of each PIN is thought to be
proportioned to its energy resolution.
Based on ground measurements using ``flight equivalent'' PIN diodes
and analog electronics,
an energy resolution of $\sim$4 keV yields a pedestal channel of $\sim$2 keV.
Therefore, the measured pedestal channel of each
PIN is assigned to an energy of 2.0 keV.
Finally, as shown in figure~\ref{fig:pin_escale},
a spline curve is derived over an energy range of 2--76 keV
for each PIN diode using the three calibration points.
The accuracy of this
scale is estimated to be about 1\%, based on deviations
of the calibration points from the spline curves.
The in-orbit energy scales agree well with the pre-launch ones
measured in an energy range of 20--50 keV, 
while some PINs show significant nonlinearities above 50 keV.

\begin{figure}
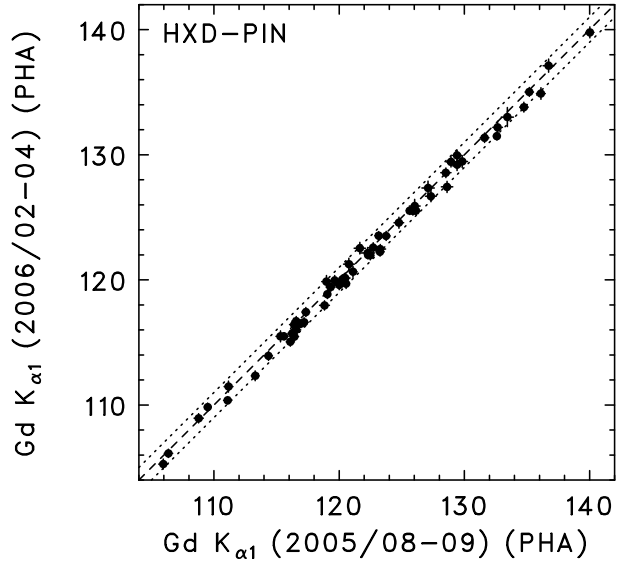

\begin{center}
\FigureFile(0.48\textwidth,0.48\textwidth){figure6.eps}
\end{center}
\caption{A comparison of the Gd-K peak channels of the 64 PINs
between two periods which are 1--2 and 7--9 months after the launch.
}
\label{fig:pin_longgain}
\end{figure}

The average counting rate of the background Gd-K line
is $\sim$ 1.2$\times$10$^{-4}$ ct s$^{-1}$ for one PIN.
Therefore, long-term variations 
of the individual gains can be monitored when accumulated over
one to two months. As shown in figure~\ref{fig:pin_longgain},
the PIN gains have stayed constant, within one ADC channel
($\sim$0.4 keV), at least for half a year.
Therefore, throughout the performance verification phase, 
each PIN diode employs a single energy scale of its own
when converting the raw ADC channels into the 
pulse height invariants (PI).

\subsection{Energy Threshold}
\label{subsection:3-2}

\begin{figure}
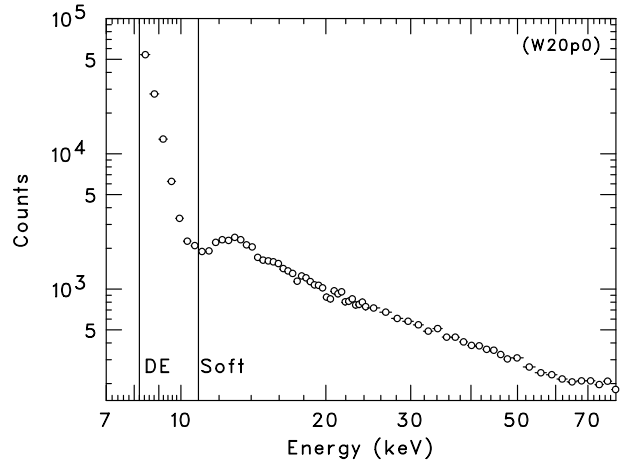

\begin{center}
\FigureFile(0.48\textwidth,0.48\textwidth){figure7.eps}
\end{center}
\caption{A typical background spectrum of one PIN. Two vertical
lines indicate the LD level applied in HXD-DE, and the energy 
threshold used in the processing software.}
\label{fig:pin_threshold}
\end{figure}

\begin{figure}
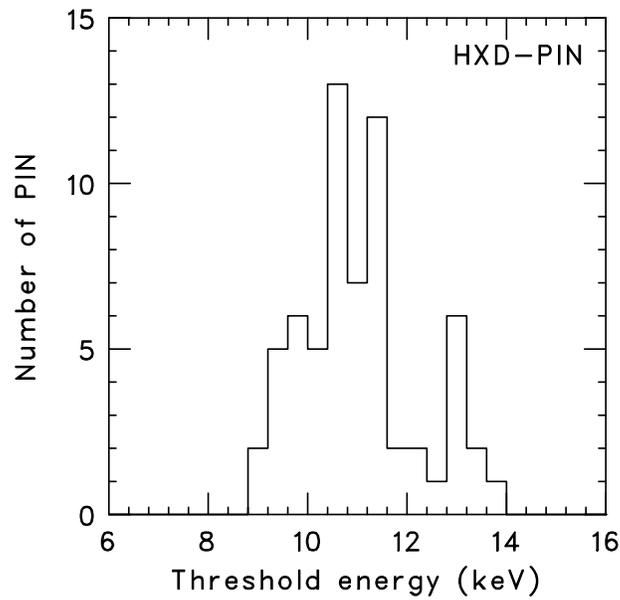

\begin{center}
\FigureFile(0.48\textwidth,0.48\textwidth){figure8.eps}
\end{center}
\caption{The number histogram of energy thresholds of the 64 PIN diodes.}
\label{fig:pin_thr_dist}
\end{figure}

As previously described in \S\ref{subsection:2-3} and 
\S\ref{subsection:2-4}, the PIN signals are already 
screened by the onboard analog and digital lower discriminator (LD) levels,
in HXD-AE and HXD-DE, respectively. However,
the PIN events transmitted to ground, as shown in figure 
\ref{fig:pin_threshold}, still contain low-energy thermal and/or 
electrical noise component, which varies significantly in orbit.
To remove these noise events, a higher threshold must be applied
by the analysis software (\S\ref{subsection:2-6}) to individual 
PIN diodes.
As shown in figure~\ref{fig:pin_threshold}, this ``software LD''
was set at the crossing point between the noise spectrum and 
non-celestial background events. A long-term stability of the
noise spectrum was also confirmed from a comparison of screened
spectra obtained at September 2005 and February 2006 (\cite{Fukazawa2006}).
Combined with the energy scale, this software LD determines the actual 
lower-limit energy of the relevant PIN diode.
Figure~\ref{fig:pin_thr_dist} shows distribution of the energy 
thresholds determined in this way.
It ranges from 9 to 14 keV, with an average of 
$\sim$10 keV, which satisfies the design goal.
After the event screening by these thresholds,
each spectrum loses its effective area below the corresponding
energy, and this effect should be correctly taken into account 
in the energy response matrix for PIN.

\subsection{Background Reduction}
\label{subsection:3-3}

\begin{figure}
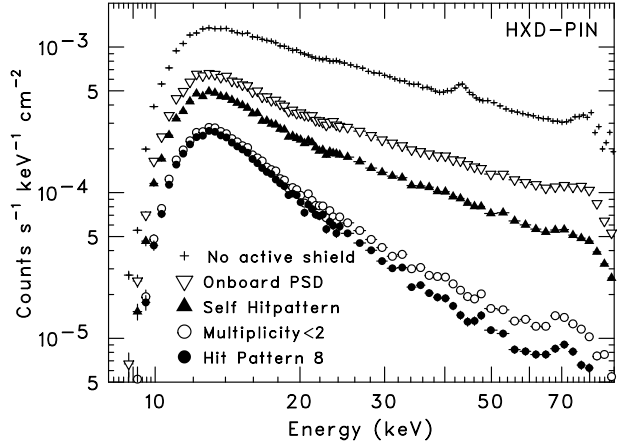

\begin{center}
\FigureFile(0.48\textwidth,0.48\textwidth){figure9.eps}
\end{center}
\caption{The background spectra summed over the 64 PINs,
acquired under various reduction conditions  (see text).
They are normalized by the total geometrical area of the 64 PIN diodes.
}
\label{fig:pin_bgdrej}
\end{figure}

After discarding the low-energy events below the individual thresholds
set by the data analysis software, 
the residual background of PIN diodes is further 
reduced by fully utilizing the anti-coincidence method, 
which comprises the basic concept of the HXD. 
Before applying the anti-coincidence,
a typical summed event rate from all the 64 PIN diodes is already 
reduced down to 2--3 ct s$^{-1}$, 
about one percent of the initial trigger rate,
most of which are caused by the in-orbit electrical interferences
or thermal noise.
Figure~\ref{fig:pin_bgdrej} illustrates how the PIN background is
reduced by stepwise application of anti-coincidence conditions.
In the figure, the crosses denote events which were extracted from 
a period when the PMT high-voltages were reduced to zero due to 
operational reasons, that is, when the BGO shields were working
only as ``passive'' shields and collimators 
rather than the active anti-coincidence counters. 
This background level is as high as those achieved in 
past hard X-ray missions equipped with passive collimators 
(\cite{Rothschild1998}; \cite{Frontera1997}).
Once the BGO shields start working and the hard-wired PSD function
is enabled (i.e., events with significant energy deposits onto
BGO are discarded in HXD-AE), the background decreases by a factor
of 3 as indicated by the open triangles in figure~\ref{fig:pin_bgdrej}. 
Since the threshold energy for the PSD is higher than that for 
the hit-pattern generation, the remaining events can be almost
halved through on-ground data screening, by discarding those events
which carry the hit-pattern flag from the same unit (filled triangles). 

The whole detector volume of HXD is always exposed 
to energetic cosmic-ray particles, 
of which the energies are higher 
than the geomagnetic cut-off rigidity (COR) of several GV, 
with a typical flux of $\sim$1 particle s$^{-1}$ cm$^{-2}$.
When they penetrate the detector, secondary radiation is 
promptly generated and adds to the background events
in surrounding units. 
Since most of the cosmic-ray particles are charged, 
their penetration usually causes simultaneous hits to multiple units. 
This ``multiplicity'', $N$, defined
by the hit-pattern signals, can be used as an efficient tool for the 
rejection of such events. Here, a valid PIN or GSO event is defined 
to have a multiplicity $N$ (0$\leq$$N$$\leq$35), if there are simultaneous
hits in $N$ units excluding the relevant triggering unit itself.
If a smaller multiplicity is required as the screening condition, 
the background will get lower,
but the signal acceptance will also decrease
due to an increase of the accidentally coinciding probability. 
With an average counting raw rate of $\sim$1 kct s$^{-1}$ in
each unit and the coincidence width of 5.6 $\mu$s, 
requiring $N\leq1$ has been found to be optimum;
that is, events are discarded if there are two or more hits 
in the hit-pattern except that from the triggering unit itself.
This condition leads to a further background reduction by 
a factor of three, as represented by open circles
in figure~\ref{fig:pin_bgdrej}.

Although applying a tighter multiplicity cut, i.e., 
requiring $N=0$ (no hit in any other unit) is unfavorable
because of a reduction in signal acceptance, 
the $N=0$ requirement works effectively 
if used under appropriately restricted conditions.
In fact, filled circles in figure~\ref{fig:pin_bgdrej}
represent the spectrum obtained by requiring $N=0$ 
in the surrounding 8 units around the triggering one. 
The detailed studies confirm that this condition (and $N\leq$1 in the remaining 27) 
optimizes the anti-coincidence condition \citep{Kitaguchi2006}.
The final background event rate obtained after applying all of 
these screening conditions is reduced to mere $\sim$0.5 ct s$^{-1}$, 
which corresponds to $\sim$3$\times$10$^{-4}$
ct s$^{-1}$ keV$^{-1}$ cm$^{-2}$ at 13 keV, 
and $\sim$1$\times$10$^{-5}$ ct s$^{-1}$ keV$^{-1}$ cm$^{-2}$ at 60 keV, 
for a geometrical area of 174 cm$^2$.

\subsection{Energy Response}
\label{subsection:3-4}

Using  Monte-Carlo simulations based on the GEANT4 toolkit
(\cite{Allison2006}; \cite{Agostinelli2003}; \cite{Terada2005}),
and implementing therein  the same event screening conditions 
as those employed by the analysis software,
energy response matrices can be constructed individually for
the 64 PIN diodes. 
Geometrical parameters, such as the size of guard-ring 
structure of PIN diodes (16.5$\times$16.5 mm$^2$; Paper I),
or individual inclinations of the fine-collimators measured
in orbit (\S\ref{subsection:6-1}), 
are precisely described in a  ``mass model'',
together with chemical composition of material.
In-orbit calibration results, such as the energy scales, 
energy resolutions, and spectral shapes set by the analog LD
which were individually measured using the Crab spectra,
are also imported into the Monte-Carlo simulation.

\begin{figure*}
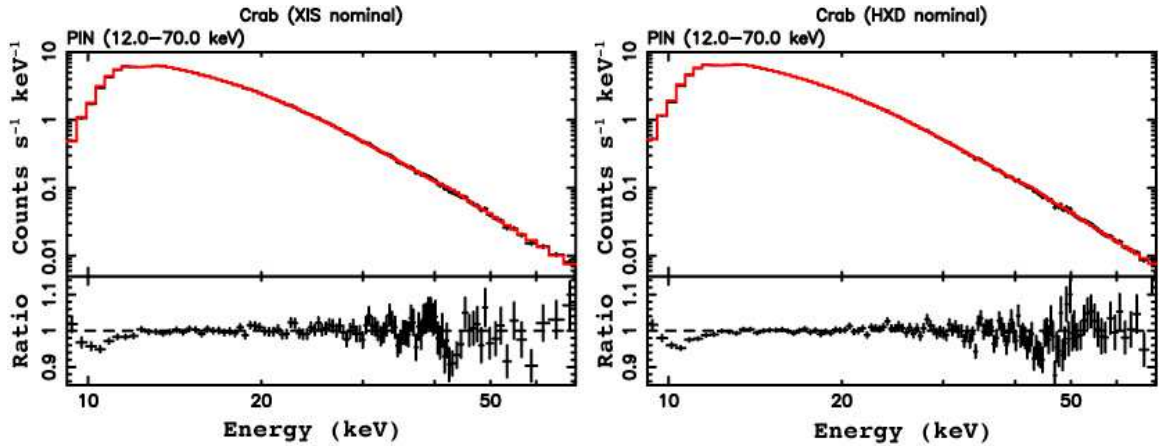

\begin{center}
\FigureFile(0.90\textwidth,0.48\textwidth){figure10.eps}
\end{center}
\caption{
The background-subtracted HXD-PIN spectra (64 summed) of the Crab nebula, 
obtained at the XIS nominal ({\it left}) and HXD nominal ({\it right}) positions, 
compared with predictions ({\it red}) by the best-fit power-law model 
of which the parameters are given in table~\ref{tbl:pin_crabfit}.
The fitting was carried out in a range of 12--70 keV, 
and then remaining channels were retrieved.
The lower panels show the data-to-model ratio.
}
\label{fig:pin_crabspec}
\end{figure*}

The silicon PIN diodes used in the HXD are so thick, $\sim$2.0 mm,
that their full depletion needs a bias voltage around 700 V (\cite{Ota1999}).
Therefore, at the nominal operation voltage of $\sim$500 V, 
the actual thickness of the depletion layer can vary among the 64 PINs, 
and should be individually calibrated.
This was done in the following two steps. 
First, the Crab spectrum of each PIN was analyzed 
for its slope in an energy range of 45--78 keV,
where the diode thickness has little effects.
The slope is determined solely by the Crab's slope and the energy
dependence of the interaction cross-section of silicon.
This analysis confirmed that the energy scales 
established in \S\ref{subsection:3-1} are correct. Then, 
the overall 15--78 keV Crab spectra were fitted individually 
by a single power-law model,
and the effective thickness was adjusted 
so that every PIN spectrum can be reproduced 
by the same photon index as obtained in the 45--78 keV band.
Finer tunings were introduced to properly model the shape of 
the efficiency decrease toward lower energies, 
which is mainly determined by the 64 analog LD levels. 
Finally, by combining the 64 response matrices into a single one,
64-PIN summed spectra can be collectively examined. 

Since Suzaku has two nominal pointing positions, (``XIS nominal''  and ``HXD nominal''), 
corresponding two response matrices were constructed
({\tt ae\_hxd\_pinxinom\_20060814.rsp} and
{\tt ae\_hxd\_pinhxnom\_20060814.rsp}).
Figure~\ref{fig:pin_crabspec} shows the Crab spectra (64 PINs summed)
measured at these two nominal positions.
There, blank-sky backgrounds,
obtained  two days before, were subtracted.
The spectra were then fitted with a single power-law model, 
using the relevant response matrix over an energy range of 12--70 keV;
the obtained best-fit parameters are summarized in table~\ref{tbl:pin_crabfit}.
Thus, the spectrum is well reproduced within a few \% over the entire range used, 
while the deviation becomes larger up to 10\% below $\sim$12 keV,
where the effective area is rapidly changing with the energy.
There is also an artificial structure at around
the characteristic X-ray energy of gadolinium, suggesting that the
modeling of the effect of active shields is yet to be improved.

\begin{longtable}{lccc}
\caption{Best-fit parameters and 90\% confidence errors for the 
PIN spectra of the Crab nebula.}
\label{tbl:pin_crabfit}
\hline\hline
Target position &  Photon index & 
Normalization\footnotemark[$*$] & $\chi_{\nu}^2$(d.o.f)\\
\endfirsthead
\hline\hline
Target position &  Photon index & 
Normalization\footnotemark[$*$] & $\chi_{\nu}^2$(d.o.f)\\
\endhead
\hline
\endfoot
\endlastfoot
\hline
XIS nominal\footnotemark[$\dagger$] & 
2.11 $\pm$ 0.01 & 11.7 $\pm$ 0.14 & 1.03 (152)\\
HXD nominal\footnotemark[$\ddagger$] & 
2.10 $\pm$ 0.01 & 11.2 $\pm$ 0.09 & 1.24 (152)\\
\hline
\multicolumn{4}{l}{\hbox to 0pt{\parbox{180mm}{\footnotesize
\par\noindent
Notes. The column density for the interstellar absorption is fixed at
3$\times$10$^{21}$ cm$^{-2}$.
\par\noindent
\footnotemark[$*$] Power-law normalization in a unit of
photons cm$^{-2}$ s$^{-1}$ keV$^{-1}$ at 1 keV.
\par\noindent
\footnotemark[$\dagger$] 
Observation performed on 2005 Sep.15 19:50--Sep.16 02:10 (UT)
\par\noindent
\footnotemark[$\ddagger$] 
Observation performed on 2006 Apr.05 12:47--Apr.06 14:13 (UT)
}\hss}}
\end{longtable}

%
%
\section{In-orbit Performance of GSO}
\label{section:4}

\subsection{Energy Scale}
\label{subsection:4-1}

Pulse heights of  the GSO events depend on  
light yields of the individual scintillators, 
the PMT gains  which are sensitive functions  of the high voltage levels,
and the amplifier settings in HXD-AE (table~\ref{tbl:aesetting}).
Among these factors,
the scintillator light yields and the PMT gains are temperature sensitive.
Before the launch, 
the GSO pulse heights in the 16 Well units  were roughly equalized 
by adjusting the high-voltage levels (per 4 Well units),
and trimming resistors in high-voltage distribution boxes.
However, the GSO pulse heights for a given energy changed across 
the launch  (\S\ref{subsection:2-3}),
and have been changing since then, 
due, e.g., to the  HXD-S temperature fluctuations (\S\ref{subsection:2-1}),
and to some unit-dependent long-term and short-term effects 
(to be detailed in \S\ref{subsection:4-3})
which are most likely taking place in the PMTs. 
The GSO pulse heights must be 
corrected for these temporal  gain changes, 
in order for them to be correctly converted to energies.
At present, the gain parameters in HXD-AE  are kept fixed
to the values reached through the initial operation (\S\ref{subsection:2-3}),
so as to keep consistency of the calibration data base.

\begin{figure}
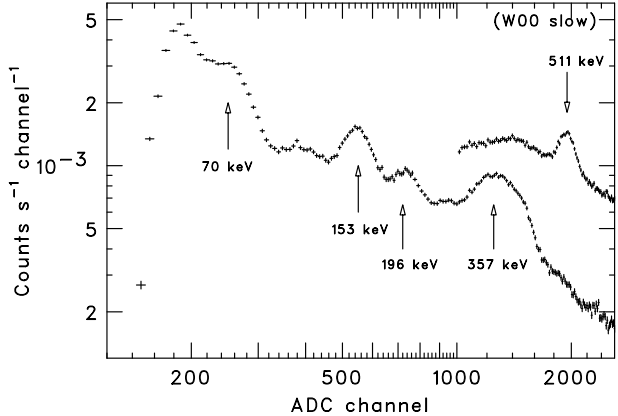

\begin{center}
\FigureFile(0.48\textwidth,0.48\textwidth){figure11.eps}
\end{center}
\caption{Typical phoswich pulse-height spectra of one Well unit, 
acquired in orbit from a blank sky.
The lower data points represent events  
obtained in the full anti-coincidence condition, 
while the higher ones, shown only above 1000 channels for clarity,
indicate those discarded by anti-coincidence.
}
\label{fig:gso_lines}
\end{figure}

The in-flight GSO energy scales 
(i.e., the relations between the incident photon 
energies and the output pulse heights),
including the temporal gain changes mentioned above,
can be determined utilizing several nuclear lines in  background spectra.
One is the broad line feature at $\sim 350$ keV 
with a count rate of $\sim$0.16 ct s$^{-1}$ per Well unit (Paper I),
produced by $\alpha$ particles from a natural 
radioactive isotope \atom{Gd}{}{152} contained in GSO.
Extensively utilized  in the pre-launch calibrations,
this broad line  also provides a good in-orbit calibrator,
because it is clearly detected in the background spectra
as shown in figure~\ref{fig:gso_lines};
under the full anti-coincidence condition (\S\ref{subsection:4-5}),
this feature can be detected in  2--3 ks of data integration.
 
In addition to the \atom{Gd}{}{152} peak, 
proton-beam irradiation experiments, 
conducted before the launch using accelerator facilities (\cite{Kokubun1999}),
predict several radioactive isotopes to be created in GSO, 
by high-energy particles including  
geomagnetically trapped protons in the SAA.
These ``activation'' isotopes, generally proton-rich,
decay via either $\beta^+$ or  electron-capture (EC) processes.
While $\beta^+$-decay species produce continuum 
and are hence useless for calibration purposes,
the EC process will give a full energy deposit in GSO,
and will produce spectral lines with well defined energies.
As shown in figure~\ref{fig:gso_lines}, 
several peaks appeared in the GSO background spectra
under the  full anti-coincidence condition  (\S\ref{subsection:4-5}),
and they are successfully identified with the corresponding isotopes 
as listed in table~\ref{tbl:calline}. 
The most prominent line,  at  $\sim$150 keV, 
is due to  EC decay of \atom{Gd}{}{153}.
Since this isotope has a half-life of 241 days,
the  line has been gradually building up after the launch
on a similar time scale (\S\ref{subsection:5-3}).
As of 2006 January, 
it has an average counting rate of
$\sim$0.05 ct s$^{-1}$  per Well,
and can be used as a second calibrator.

The proton-induced activation takes place not only in GSO,
but also in the surrounding BGO shields.
Subsequent  $\beta^{+}$-decay  events in BGO
are usually detected as  ``multi-hit''  events,
because each of them produces a pair of  annihilation photons at 511 keV
which are generally  detected by  neighboring units.
Although these events are rejected in 
normal observations by anti-coincidence,
the 511 keV peak is clearly detected, 
as shown in figure~\ref{fig:gso_lines}, 
if such {\em multi-hit} events are purposely accumulated.
By fitting these pulse-height spectra, 
in-orbit energy resolutions of individual Well units,
at $\sim$500 keV, were measured and confirmed to be almost 
the same as those obtained on-ground calibrations ($\sim$11\%; Paper I).
The  annihilation line intensity decreases rapidly after each SAA passage, 
on a typical time scale of $\sim$10 ks, 
resulting in a day-averaged rate of $\sim$0.1 ct s$^{-1}$ per Well unit. 

Thus, the in-orbit data contains at least three peaks
with secure energy identifications;
511 keV, 350 keV, and 153 keV.
By fitting them with a linear function, 
the first approximation to the GSO energy scales of 
the individual Well units were obtained as
\begin{equation}
 P = a_i E - b_i ~~.
 \label{eq:gso_linear}
\end{equation}
Here,  $E$ is the photon energy in keV, 
$P$ is the pulse height in units of ADC channels,
$a_i \sim$4 channel keV$^{-1}$ 
and $b_i \sim$50 channels are positive parameters,
and $i=0,1,..15$ is the unit number.

\begin{longtable}{llll}
\caption{Activation lines used for in-orbit calibration of the GSO energy scale.}
\label{tbl:calline}
\hline\hline
Energy (keV)  & Radioactive isotope & Decay mode & Half life\\
\endfirsthead
\hline\hline
Energy (keV)  & Radioactive isotope & Decay mode & Half life\\
\endhead
\hline
\endfoot
\endlastfoot
\hline
70.0  & \atom{Gd}{}{151} & EC & 124 (day)\\
153.0 & \atom{Gd}{}{153} & EC & 241 (day)\\
196.0 & \atom{Eu}{}{151m} & IT & 60 ($\mu$s) \\
$\sim$357   & \atom{Gd}{}{152}(natural) & $\alpha$ & 1.1$\times$10$^{14}$ (year)\\
511.0 & various & $\beta^{+}$ & \\
\hline
\end{longtable}

\subsection{Corrections for Nonlinearity}
\label{subsection:4-2}

The offsets $-b_i <0$ in equation (\ref{eq:gso_linear}) are due to nonlinearities, 
both in the GSO light yields and the HXD-AE performance, 
which are known to become significant  in energies below $\sim 100$ keV
(\cite{Uchiyama2001}; \cite{Kawaharada2004}; Paper I). 
Therefore, equation  (\ref{eq:gso_linear}) is thought to be accurate
only in energies above 150 keV.
If, in fact,  equation (\ref{eq:gso_linear}) is  tentatively 
extrapolated to below 150 keV, 
it underestimates the pulse height 
of another  emission line seen in the data,
namely the 70 keV line originating from 
EC-decays of \atom{Gd}{}{151} (table~\ref{tbl:calline}).
As a result, all Well units were reconfirmed to show
the expected nonlinearities with the same sense as
indicated by the negative offset $-b_i$,
but with larger deviations ($\sim$10\% at 70 keV) 
than was measured in the on-ground calibrations ($\sim$3\%).

In order to represent the nonlinearity,
equation (\ref{eq:gso_linear}) has been improved empirically as
\begin{equation}
 P = a_i E - b_i + c_i \exp(-E/d_i) ~~,
\label{eq:gso_nonlinear}
\end{equation}
where $c_i \sim$280 channels and $d_i \sim$35 keV  
are additional positive parameters.
The values of $c_i$ and $d_i$ were determined 
using the activation peak at 70 keV 
and K-edge energy of gadolinium (50 keV; \S\ref{subsection:5-3}), 
and also requiring the pulse heights for  $E=0$ (namely $P= -b_i+c_i$ channels)
to coincide with the ``pedestal'' ADC channels  (\S\ref{subsection:3-1})
which are measured in orbit by inspecting events triggered 
by any one of the four PIN diodes.

\begin{figure}
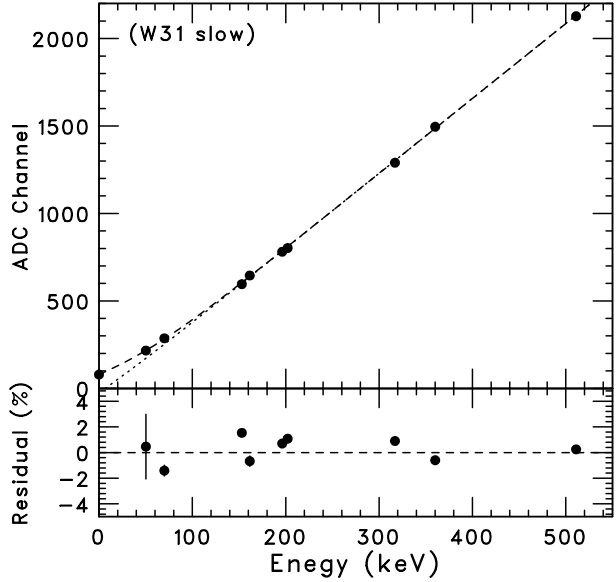

\begin{center}
\FigureFile(0.48\textwidth,0.48\textwidth){figure12.eps}
\end{center}
\caption{Energy scales of GSO in a particular Well unit,
determined with equation (\ref{eq:gso_linear})(dotted line) and 
equation (\ref{eq:gso_nonlinear})(dashed curve),
fitted to various energy-scale calibration features.
The lower panel shows residuals from the dashed curve.
}
\label{fig:gso_nonlinear}
\end{figure}

As shown in figure~\ref{fig:gso_nonlinear}, 
the empirical energy scales of equation (\ref{eq:gso_nonlinear})
have been confirmed to account for the employed 
calibration features within $\pm$3\%, in every unit, 
over the entire energy range from 50 to 600 keV.
An independent confirmation of equation (\ref{eq:gso_nonlinear})
has been  obtained from comparisons
between the PIN and GSO spectra 
of two X-ray pulsars with cyclotron absorption lines 
at around 35--45 keV (\cite{Terada2006}).

\subsection{Temporal Gain Changes}
\label{subsection:4-3}

\begin{figure}
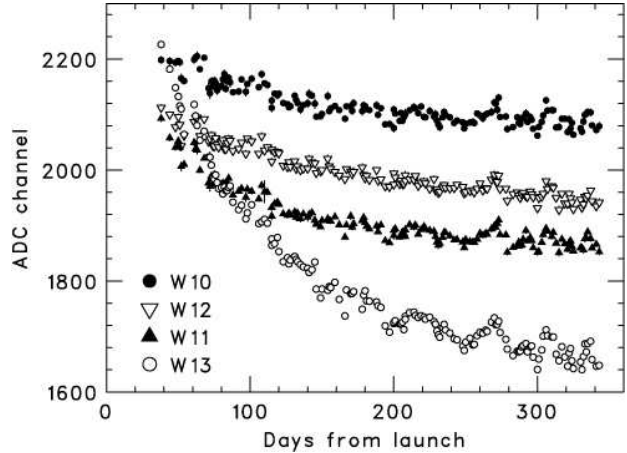

\begin{center}
\FigureFile(0.48\textwidth,0.48\textwidth){figure13.eps}
\end{center}
\caption{Long-term variations of the GSO gains in four Well units
which are biased by the same high-voltage supplier,
determined in reference to the 511 keV line,
from 2005 September to 2006 May.
}
\label{fig:gso_gainhist}
\end{figure}

Figure  \ref{fig:gso_gainhist} shows long-term GSO 
gain histories of several representative Well units,
determined referring to the pulse heights of the 511 keV line
which has the highest signal-to-noise ratio among the available calibrators.
Each data point in these histories represents the average over an observation,
which lasts typically one to two days.
Over the first 9 months in orbit,
all the Well  units have thus exhibited gradual gain decreases, 
by  5\% (minimum) to 20\% (maximum).
In addition to the long-term trends,
figure  \ref{fig:gso_gainhist} also reveals short-term gain fluctuations,
which are anti-correlated with the temperature history 
of figure~\ref{fig:temp_history}.
This is because the GSO light yield and the PMT gain 
both depend inversely on the temperature,
with typical coefficients of 
$\sim 0.5\%$ C$^{-1}$ and $\sim$0.2\% C$^{-1}$, respectively.

\begin{figure}
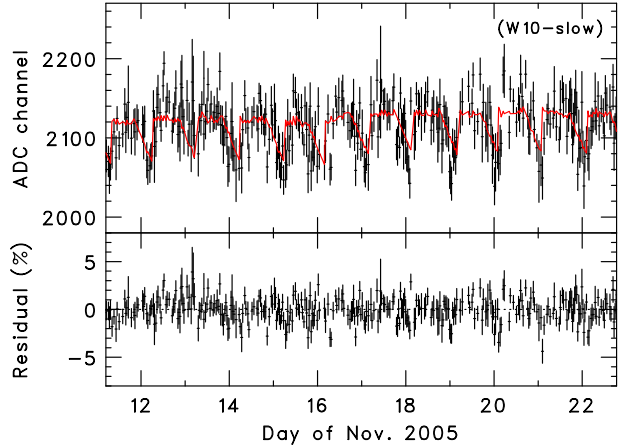

\begin{center}
\FigureFile(0.48\textwidth,0.48\textwidth){figure14.eps}
\end{center}
\caption{Short-term GSO gain variations in a representative Well unit, 
measured every 2 ks  during $\sim$10 days in 2005 November.
The top panel shows peak channels of the 511 keV line
and the prediction of an empirical model ({\it red}), 
while the bottom panel shows deviations from the model.
}
\label{fig:gso_ghist_short}
\end{figure}

Figure~\ref{fig:gso_ghist_short} represents 
a short-term gain history of a particular Well unit,
constructed by accumulating its GSO events 
into a series of spectra every 2 ks,
and fitting the 511 keV line therein.
It reveals a periodic  variation by a few percent
on a yet shorter time scale than that of the temperature fluctuation,
synchronized with  the orbital revolution of Suzaku ($\sim$96 min).
This is because the PMT gains slightly jump up
when the high voltages are resumed after spending the off period
in the SAA, 
and then gradually decrease in somewhat unit dependent ways.
Therefore, the gains show almost stable levels in ``SAA orbits''
which contain the high-voltage resets, whereas they continue to 
decline in ``non-SAA orbits'' until the next reentry.
This behavior is modeled by an empirical function,
in terms of the measured temperature of HXD-S
and the time after high-voltage resumption (\cite{Fukazawa2006}).
The model parameters are adjusted,
unit by unit, by fitting the actual gain histories.
As exemplified by red curves in  figure~\ref{fig:gso_ghist_short},
the model can reproduce the instantaneous GSO gains,
namely the parameter $a_i$ in  equation (\ref{eq:gso_nonlinear}),
to an accuracy of  3\% as a function of time
with a typical time resolution of 2 ks.

\subsection{Background Reduction with PSD}
\label{subsection:4-4}

As described in \S\ref{subsection:2-4},
the HXD events are normally transmitted to ground
after screened first by the hard-wired PSD in HXD-AE,
and then by the digital PSD in HXD-DE.
Nevertheless, the data still contain significant background events,
namely residual BGO events at lower energies ($\lesssim 100$ keV),
and residual Compton events at higher energies.
In order to achieve the highest sensitivity to celestial signals,
the PSD criteria must be further tightened and optimized
in off-line data analyses,
by discarding these background events as much as possible
but retaining the signal acceptance.
This process is hereafter called ``software PSD''.

\begin{figure}
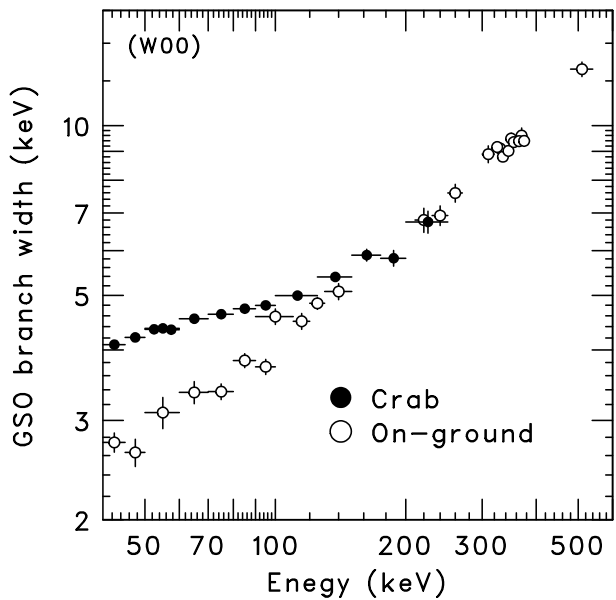

\begin{center}
\FigureFile(0.48\textwidth,0.48\textwidth){figure15.eps}
\end{center}
\caption{The GSO branch width (Gaussian sigma) on the fast-slow diagram,
shown as a function of energy.
Open black circles indicate pre-launch measurements using isotopes,
while filled circles refer to the Crab Nebula data after background subtraction.
}
\label{fig:gso_psd_sigma}
\end{figure}

\begin{figure}
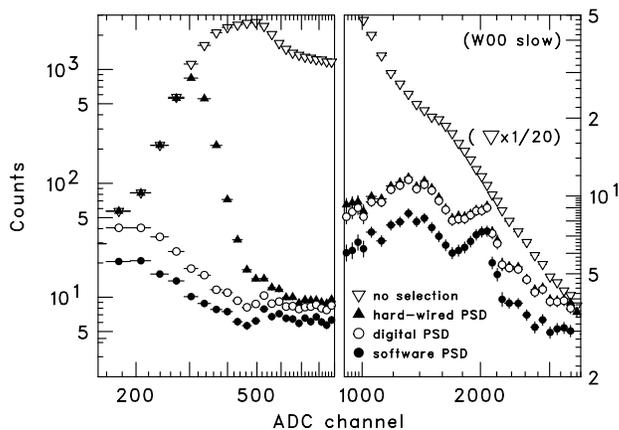

\begin{center}
\FigureFile(0.48\textwidth,0.48\textwidth){figure16.eps}
\end{center}
\caption{Raw in-orbit background spectrum of a Well unit without
any PSD selection, compared with those after applying the hard-wired,
digital, and software PSD selections.
Data were  integrated during the daily diagnostic runs,
when the onboard PSD function is temporarily switched off
(\S\ref{subsection:2-5}). 
The spectra are shown separately below and above 900 channel
in the left and right panels, respectively.
For clarity, the raw background spectrum  in the right panel
is reduced in intensity by a factor of 20.
}
\label{fig:gso_psdthr_spec}
\end{figure}

In order to find the optimum software PSD condition,
\citet{Kitaguchi2006} analyzed the two-dimensional
fast-slow diagram (Paper I) of the  Crab Nebula,
unit by unit, after subtracting blank-sky backgrounds.
They  quantified Gaussian-equivalent ``spread''  
$\sigma(E)$ of the GSO events on the diagram,
in the direction ($-45^{\circ}$)
which is perpendicular to its branch.
As shown in figure~\ref{fig:gso_psd_sigma}, 
the obtained width of GSO branch, 
as a function of the energy, 
shows a good agreement with those measured on ground above 150 keV, 
while slightly broadens at lower energy range probably due to the
in-orbit nonlinearity effect of the PMTs (\S\ref{subsection:4-2}).
Then, the Crab Nebula data and the blank-sky background
were screened in the same manner,
using  $\pm x \sigma(E)$ cuts 
where $x>0$ is the cut condition  to be optimized.
By examining how the residual blank-sky background
and the Crab signals change with $x$,
it was found that $x=2.1$,
which corresponds to a signal acceptance of 96\%,
generally maximizes the signal-to-noise ratio
under background-dominant conditions.
Through this optimum software PSD cut,
the GSO background data transmitted to ground 
has  been further reduced by 50\% 
as illustrated in  figure \ref{fig:gso_psdthr_spec}.

\subsection{Background Reduction with the Anti-Coincidence}
\label{subsection:4-5}

After applying the software PSD within each Well unit,
the GSO background, like the PIN data, 
can be further reduced using the anti-coincidence function
working among multiple Well and Anti units.
This screening is  particularly useful in eliminating
secondary radiation produced by cosmic-ray charged particles,
and the 511 keV lines originating from $\beta^{+}$ decays 
of activated nuclei in the BGO shield (\S \ref{subsection:4-1}).
However, like the PSD case,
too severe anti-coincidence conditions would reduce  
the signal acceptance via chance coincidence;
accordingly, the conditions need to be optimized.

\begin{figure}
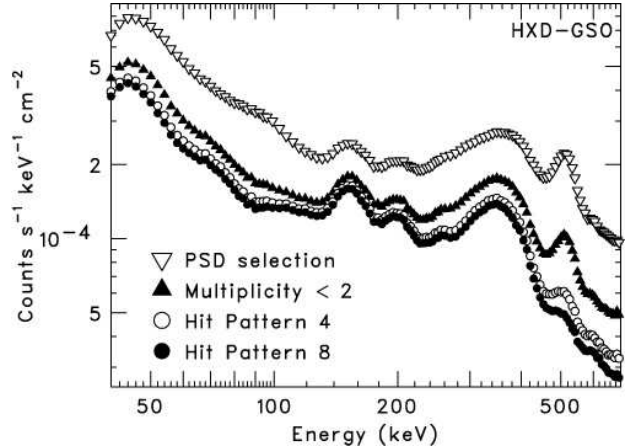

\begin{center}
\FigureFile(0.48\textwidth,0.48\textwidth){figure17.eps}
\end{center}
\caption{The background spectra summed over the 16 Well units,
acquired under various anti-coincidence conditions  (see text).
They are normalized by the total geometrical area of the 64 GSO scintillators.
}

\label{fig:gso_bgdrej_hp}
\end{figure}

As detailed in  \citet{Kitaguchi2006},
the optimum anti-coincidence condition has been found to 
reject  a GSO event (surviving the full PSD cut) from a Well unit
as backgrounds,
if either of the following two conditions are satisfied; 
\begin{enumerate}
\item
It has a simultaneous hit in at least one of the 8 units
that surround the relevant Well.
\item
It has at least two simultaneous hits in any 
units other than the relevant unit itself.
\end{enumerate}
As shown in figure~\ref{fig:gso_bgdrej_hp},
applying this condition has nearly halved the GSO background
over the entire  50--500 keV range.
In particular, the 511 keV lines have been
reduced to a level of $\sim$0.05 ct s$^{-1}$,
because most of them are emitted in pairs from $\beta^{+}$-decay nuclei,
and hence they produce double hits.
The measured in-orbit SLD (slow-LD; \S\ref{subsection:2-3}) 
rate is  1000$\sim$1500 ct s$^{-1}$ per Well,
implying a chance probability of  4--6\% 
for the surrounding 8 units to cause an accidental rejection
of a valid GSO signal.

\subsection{Energy Response}
\label{subsection:4-6}

In the same way as the PIN diodes (\S \ref{subsection:3-4}),
the GEANT4 Monte-Carlo toolkit was utilized to construct
the GSO energy response function.
This technique is particularly important 
in energies above $\sim 100$ keV,
where the probability of signal photons 
undergoing Compton scattering increases,
and hence off-diagonal elements in the response matrix becomes significant.
Their analytic calculation would be difficult,
because it must take into account the  probability of a Compton-scattered  
signal photon  to be rejected  either by the PSD within the same Well unit,
or by  the anti-coincidence with another unit.
The Monte-Carlo calculations were performed 
employing the basic interaction cross sections
and detailed detector geometry,
as well as light yields of the three scintillator components
(GSO, BGO bottom piece, and BGO top piece)
measured in pre-launch tests.

\begin{figure*}
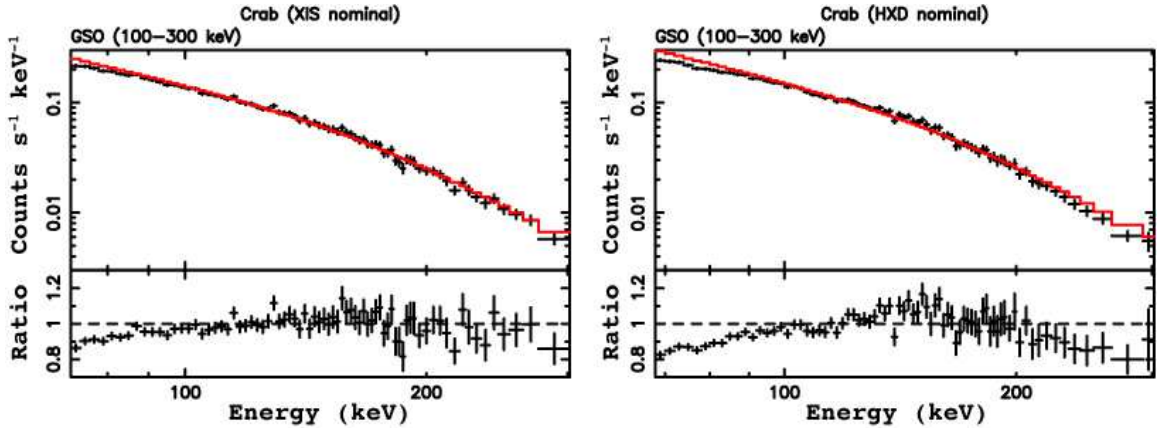

\begin{center}
\FigureFile(0.90\textwidth,0.48\textwidth){figure18.eps}
\end{center}
\caption{The background-subtracted Crab spectrum of GSO
(summed over the 16 Well units), obtained at the XIS nominal 
({\it left}) and HXD nominal ({\it right}) positions,
compared with predictions ({\it red}) by the best fit power-law model
of which the parameters are given in table \ref{tbl:gso_crabfit}. 
The fittings were carried out in a range of 100--300 keV, 
and then remaining channels were retrieved. 
The lower panels show the data-to-model ratio.
}
\label{fig:gso_crabspec}
\end{figure*}

Figure~\ref{fig:gso_crabspec} shows the GSO spectra
(summed over the 16 units) of the Crab Nebula,
after subtracting blank-sky background taken on the next day.
The backgrounds were extracted using the same ``good time interval''
conditions as the on-source data, that is, 
COR $> 8$, elapsed time from the SAA $> 500$ sec, 
and elevation from the earth rim $> 5^{\circ}$.
The spectrum is fitted with a single power-law model,
using the response function as constructed above
({\tt ae\_hxd\_gsoxinom\_20060321.rmf} and 
{\tt ae\_hxd\_gsohxnom\_20060321.rmf}).
The fit is performed over the 100--300 keV range,
to avoid poor signal-to-noise ratio above 300 keV
(where the signal becomes $\sim 20\%$ of the background)
and insufficient tuning of the PSD efficiency and other parameters below 100 keV.
Thus, the data are successfully reproduced to within $\pm 10$\%,
and the obtained best-fit parameters, 
given in  table~\ref{tbl:gso_crabfit},
agree within $\sim$10\% with those determined with PIN 
(table~\ref{tbl:pin_crabfit}).
The remaining task is to improve the fit to the Crab spectrum,
and to extend the fit toward lower and higher energies.

\begin{longtable}{lccc}
\caption{Best-fit parameters and 90\% confidence errors for the GSO 
spectra of the Crab Nebula at the XIS and HXD nominal positions.}
\label{tbl:gso_crabfit}
\hline\hline
Target position &  Photon index & 
Normalization\footnotemark[$*$] & $\chi_{\nu}^2$(d.o.f)\\
\endfirsthead
\hline\hline
Target position &  Photon index & 
Normalization\footnotemark[$*$] & $\chi_{\nu}^2$(d.o.f)\\
\endhead
\hline
\endfoot
\endlastfoot
\hline
XIS nominal\footnotemark[$\dagger$] & 
2.12 $\pm$ 0.03 & 10.6 $\pm$ 1.4 & 1.07 (98)\\
HXD nominal\footnotemark[$\ddagger$] & 
2.15 $\pm$ 0.03 & 11.7 $\pm$ 1.4 & 1.50 (96)\\
\hline
\multicolumn{4}{l}{\hbox to 0pt{\parbox{180mm}{\footnotesize
\par\noindent
Notes. The column density for the interstellar absorption is fixed at
3$\times$10$^{21}$ cm$^{-2}$.
\par\noindent
\footnotemark[$*$] Power-law normalization in a unit of
photons cm$^{-2}$ s$^{-1}$ keV$^{-1}$ at 1 keV.
\par\noindent
\footnotemark[$\dagger$] 
Observation performed on 2006 Apr.04 02:55--14:20 (UT)
\par\noindent
\footnotemark[$\ddagger$] 
Observation performed on 2006 Apr.05 12:47--Apr.06 14:13 (UT)
}\hss}}
\end{longtable}

%
%
\section{Non X-ray Background}
\label{subsection:5}

The HXD is designed to achieve a high signal-to-noise ratio
by reducing the detector background as much as possible (Paper I).
Although the HXD has no capability of rocking on-off observations
utilized in the PDS onboard BeppoSAX  and the HEXTE onboard RXTE, 
a high sensitivity can be obtained by subtracting 
a sufficiently accurate  ``modeled''  non X-ray background (NXB)
instead of the off-observation spectrum.
Given the accuracy of in-orbit calibrations,
the performance of the experiment solely depends on the reproducibility of the NXB modeling, 
and hence on the precise knowledge of temporal and spectral NXB variations.
In the near-Earth orbit of Suzaku,
the HXD field-of-view is blocked periodically  by the Earth
for a certain fraction of each orbital revolution, 
and hence actual in-orbit behavior of the NXB can 
be constantly monitored using this ``earth occultation'' data.

\subsection{Common Properties of the PIN/GSO background}
\label{subsection:5-1}

Generally, the NXB of a hard X-ray instrument,  flown in a low Earth orbit, 
consists of several components as follows:
a) delayed emissions from radio-active isotopes induced inside the detector 
mainly by SAA protons via nuclear interactions,
b) prompt secondary radiation caused by interactions between cosmic-ray 
particles and the spacecraft, and
c) intrinsic background caused by natural radioactive isotopes
in detector materials.
While the third component is constant throughout the mission,
the first and second ones significantly vary in the orbit, 
corresponding to individual depths of the SAA passages 
and/or changes in the cosmic-ray fluxes.
Therefore, the primary information about the nature of NXB can be obtained by 
measuring time variations and geographical distributions of such high energy particles.
For this purpose, the counting rates of PIN-UD from the 16 Well units
can be used as a real-time flux monitor, 
even during the SAA passages (\S\ref{subsection:2-3}), 
by regarding PIN-UD counts per a fixed period 
as the number of energetic charged particles 
which  penetrated the BGO shields and the passed  through the PIN diodes.

\begin{figure}
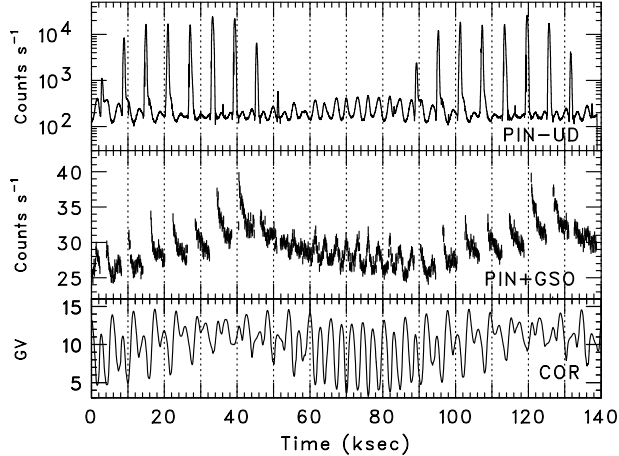

\begin{center}
\FigureFile(0.48\textwidth,0.48\textwidth){figure19.eps}
\end{center}
\caption{Typical light curves of the PIN-UD counting rate summed 
over the 16 units ({\it top}), ``cleaned'' events from PIN 
and GSO ({\it middle}), and the cut-off rigidity ({\it bottom}), 
obtained from $\sim$1.5 days observation of a blank sky field.}
\label{fig:pinud_lc}
\end{figure}

Figure~\ref{fig:pinud_lc} shows typical light curves of 
the total PIN-UD counting rate, the event rate of PIN and GSO 
after the all screening procedures have been applied, 
and the value of COR along the spacecraft orbit.
Since these were obtained from a blank-sky  observation, 
the event rate can be roughly regarded as that of the NXB. 
In the top panel, sharp peaks reaching $\sim$20000 ct s$^{-1}$, 
which corresponds to a flux of $\sim$100 ct s$^{-1}$ cm$^{-2}$,
indicate SAA passages. 
Although this flux is about an order of magnitude lower 
than the well known SAA flux at an inclination of 32$^{\circ}$ (\cite{Zombeck}), 
this is because the PIN diodes are embedded in the thick BGO shields 
and hence only SAA protons above $\sim$100 MeV reaches PIN.
Outside the SAA, 
the counting rate is also  modulated from $\sim$100 to $\sim$400 ct s$^{-1}$ 
with a period of $\sim$3000 s.
Its  clear anti-correlation with  the COR value means
that the cut-off energy of cosmic-ray particles decreases
when the satellite passes through high latitudinal regions.
The PIN-UD counts  integrated for a day are typically 
$\sim$7$\times$10$^7$, and roughly 90\% of them is the SAA protons.

The component ``a)'' mentioned above usually shows  complicated variations, 
since it is a composite of many radioactive isotopes with different half lives.
As shown in the middle panel of figure~\ref{fig:pinud_lc}, ``short-nuclides'', 
which have  half-lives shorter than the orbital period ($\sim$100 min), 
cause a rapid decrease  in the light curve after every SAA passage,
while ``middle-nuclides'', 
whose decay  time constants are longer than the revolution but shorter than a day, 
produce  a gradual increase and a decline over the SAA and non-SAA orbits, 
respectively. 
In addition, ``long-nuclides'', 
which  have life times of a few days to more than a hundred days, 
are gradually accumulated,
until they individually achieve 
equilibria between the decay and production, 
and contribute as a constant component in the light curve.

\begin{figure}
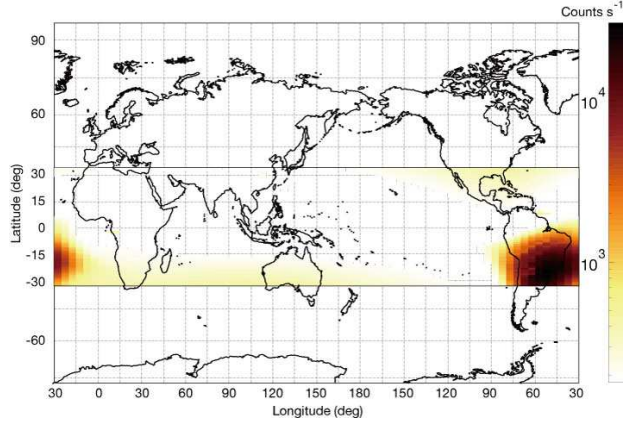

\begin{center}
\FigureFile(0.48\textwidth,0.48\textwidth){figure20.eps}
\end{center}
\caption{The flux map of cosmic-ray and trapped 
particles measured by the PIN-UD count.}
\label{fig:pinud_map}
\end{figure}

By accumulating the PIN-UD counting rate at a given position of the satellite,
and projecting the value onto the corresponding geographical coordinates, 
a flux map of high energy particles is obtained as shown in figure~\ref{fig:pinud_map}.
It was confirmed that the SAA has its centroid at around
(320$^{\circ}$, -30$^{\circ}$) with a size of 
$\sim$60$^{\circ}$$\times$40$^{\circ}$,
at the altitude of Suzaku ($\sim$570 km).
This information is being utilized to generate the high-voltage reduction 
commands in daily operations (\S\ref{subsection:2-2}).
Since the SAA is known to move  westward slowly ($\sim$0.3$^{\circ}$ per year),
and the flux of trapped protons is affected by the Solar activity, 
these maps obtained every 50 days were examined for
possible temporal changes of the position, size, and intensity of the SAA.
Then, the SAA has been confirmed to be quite stable throughout 
the performance verification phase.

\begin{figure}
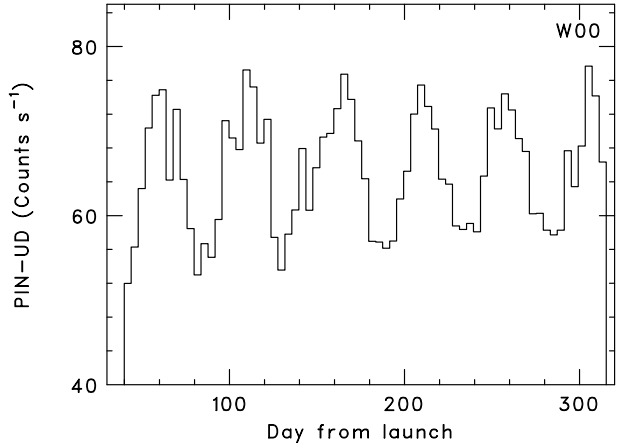

\begin{center}
\FigureFile(0.48\textwidth,0.48\textwidth){figure21.eps}
\end{center}
\caption{A light curve of the daily averaged PIN-UD count
from a representative Well unit.}
\label{fig:pinud_cycle}
\end{figure}

Even though the properties of the SAA remaine unchanged, 
the daily integrated PIN-UD counts change day by day, 
mainly due to  different total dose 
caused by different sets of SAA penetration trails of the spacecraft, 
and to a lesser extent,
due to a difference in the  satellite attitude as it gets into the SAA.
Due to an orbital precession by  $\sim 7^\circ.2$ per day,
 positions of the Suzaku's 15 daily revolutions relative to the SAA
change with a period of $\sim 50$ d.
As shown in figure~\ref{fig:pinud_cycle}, 
the daily integrated PIN-UD count indeed shows 
cyclic variations  with the same period. 
The most variable units located at the corner of the 16 Well units
exhibit a peak-to-peak variation amplitude of $\sim$50\%, 
while those of the heavily shielded central four units are  smaller than 10\%.
Although the proton flux in the SAA itself varies 
by a factor of $\sim$2 according to the 11-year solar cycle, 
such a long-term effect is not yet observable within the first  year of Suzaku.

\subsection{Properties of the Residual Background of PIN}
\label{subsection:5-2}

\begin{figure}
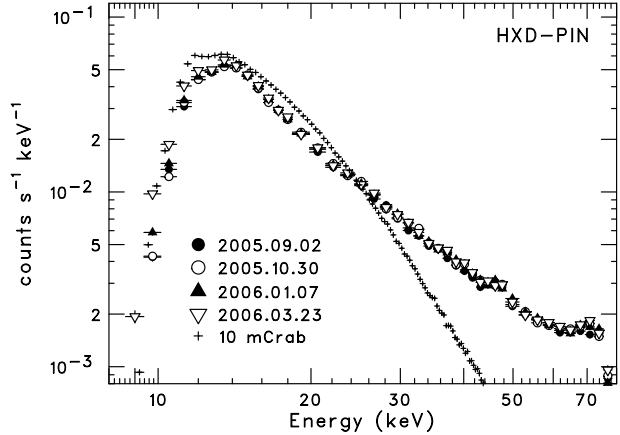

\begin{center}
\FigureFile(0.48\textwidth,0.48\textwidth){figure22.eps}
\end{center}
\caption{A comparison of four average NXB spectra measured by HXD-PIN,
on 4 occasions separated by two months.
Each observation has an exposure longer than two days.
The Crab spectrum, scaled down by two orders of magnitude, are also shown.}
\label{fig:pin_bgd_long}
\end{figure}

\begin{figure*}
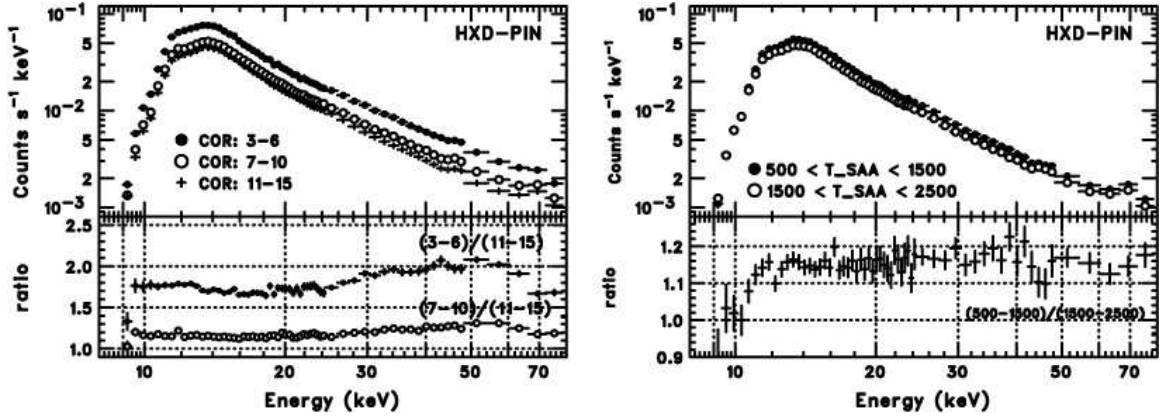

\begin{center}
\FigureFile(0.90\textwidth,0.48\textwidth){figure24.eps}
\end{center}
\caption{The PIN-NXB spectra sorted with respect to COR ({\it left}) 
and T\_SAA ({\it right}), extracted from the earth occultation data. 
Their  ratios  are  shown in bottom panels.}
\label{fig:pin_bgd_spec}
\end{figure*}

As sketched in \S\ref{subsection:5-1},
the presence  of  {\it long-nuclides} would significantly 
complicate the background reproducibility.
This is however not expected to be the case with PIN,
since long-lived radioactive isotopes are rare among 
elements  with small atomic numbers like silicon.
Figure~\ref{fig:pin_bgd_spec} compares 
four NXB spectra of PIN (hererafter PIN-NXB),
obtained  every two months over half a year.
Although blank-sky data contain both 
cosmic X-ray background (CXB) and NXB, 
the former is only $\sim$5\% of the latter 
due to the narrow field-of-view ($\sim34'$) of PIN, 
and hence the CXB sky fluctuation can be ignored.
Above 15 keV, the average PIN-NXB has thus stayed constant within 5\%, 
confirming that the PIN-NXB is free from any long-term accumulation.
When compared with a {\it scaled} Crab spectrum
(also shown in figure~\ref{fig:pin_bgd_long}),
the average PIN-NXB is roughly equivalent to a 10 mCrab source below 30 keV.
This in turn means that a sensitivity of 0.5 mCrab can be achieved 
by modeling the PIN-NXB spectrum with an accuracy of 5\%.
Given the absence of {\it long-nuclides} in the PIN diodes,
its background reproducibility is set solely by the accuracy
with which the short-term (less than a day) variations can be modeled.

\begin{figure}
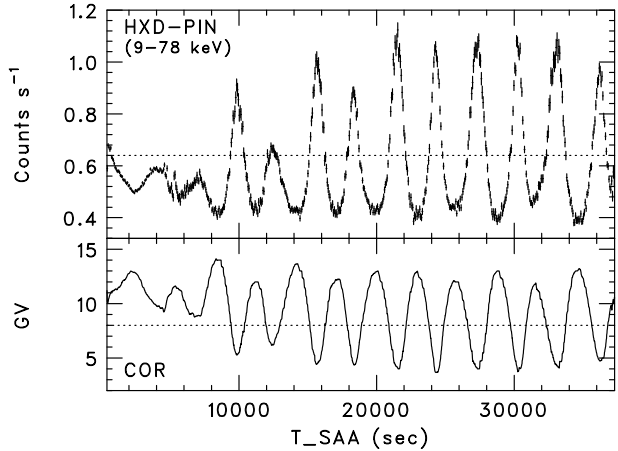

\begin{center}
\FigureFile(0.48\textwidth,0.48\textwidth){figure23.eps}
\end{center}
\caption{A light curve of the PIN-NXB folded with T\_SAA in an energy 
range of 9-78 keV ({\it top}), and a plot of cut-off rigidity obtained
as an average for each T\_SAA ({\it bottom}).}
\label{fig:pin_bgdlc}
\end{figure}

Although the anti-coincidence system of the HXD efficiently works
to veto cosmic-ray events in which charged particles interact with the detector, 
short-term behavior of NXB is still affected by secondary emissions 
from interactions between cosmic-ray particles and the spacecraft.
In addition, {\it short-} and {\it middle-nuclides} induced during 
SAA passages can also contribute to the NXB variation. 
Top panel of figure~\ref{fig:pin_bgdlc} shows a light curve of 
PIN-NXB folded with an ``elapsed time from the SAA (T\_SAA)'',
which is reset to zero at every entry to the SAA. 
Since only data during the earth occultations are utilized, 
the CXB is not included in this case. 
The bottom panel of figure~\ref{fig:pin_bgdlc} show
the averaged COR corresponding to each T\_SAA,
where clear modulation appears 
because COR and T\_SAA are mutually coupled.
A strong anti-correlation between the PIN-NXB and COR is evident, 
whereas the  dependence on T\_SAA itself is rather weak.
Therefore, the PIN-NXB is dominated by the cosmic-ray component,
rather than by the SAA components. 
The  peak-to-peak amplitude of the variation reaches 
a factor of three, but it is significantly reduced to $\sim$1.5 
when a selection criterion of COR$>$8 is applied,
as indicated by dotted lines in figure~\ref{fig:pin_bgdlc}. 
With this condition, 
the average PIN-NXB counting rate is $\sim$0.5 ct s$^{-1}$
over the entire energy range. 
It in turn means that 
statistical fluctuations of the integrated PIN-NXB counts 
become smaller than $\sim$1\% 
when the exposure exceeds 20 ks.

To construct precise models of the PIN-NXB, it is crucial to 
examine spectral variations as a function of  the COR and  T\_SAA parameters.
The left panel of figure~\ref{fig:pin_bgd_spec} shows the PIN-NXB
spectra extracted from COR regions of 3--6, 7--10, and 11--15 GV, 
together with their ratios. 
When the COR decreases, 
the PIN-NXB spectrum thus keeps a similar spectral shape below $\sim$25 keV, 
whereas it shows a significant ``hardening'' above that energy. 
On the other hand, as shown in the right panelof figure~\ref{fig:pin_bgd_spec}, 
the spectral shape of PIN-NXB  depends little on T\_SAA,
except in the lowermost energies below 12 keV.

\begin{figure}
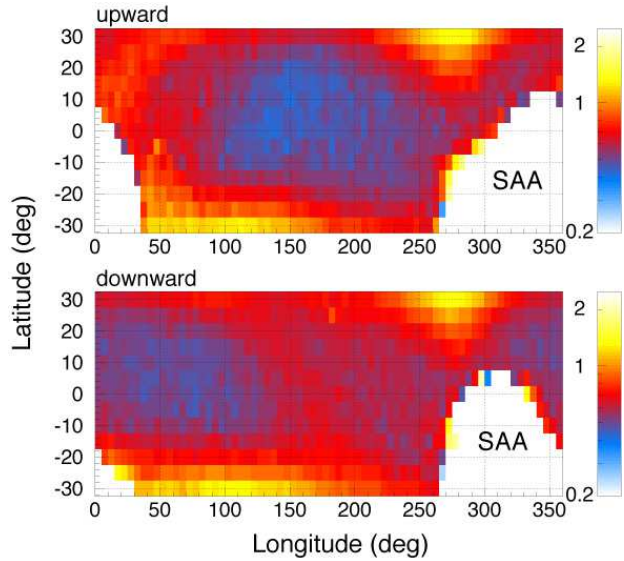

\begin{center}
\FigureFile(0.48\textwidth,0.48\textwidth){figure25.eps}
\end{center}
\caption{Maps of the total counting rate of PIN-NXB from the
entire energy range, plotted on geographical longitudes and latitudes,
when the satellite is moving toward north ({\it top}) and south ({\it bottom}).}
\label{fig:pin_bgdmap}
\end{figure}

The above parameterization of PIN-NXB  in terms of COR and T\_SAA
may be improved by replacing them
with geographical longitudes and latitudes, 
because COR and T\_SAA  are not actually independent, 
and because the T\_SAA parameter suffers a systematic 
uncertainty caused by the definition of SAA boundary.
As shown in figure~\ref{fig:pin_bgdmap}, 
geographical maps of PIN-NXB can be obtained by accumulating the counts
as a function of the instantaneous spacecraft position on the Earth. 
Two different maps are obtained according to the directions of  the satellite motion, 
namely, north-east and south-east. 
They must be distinguished,
because the elapsed time from the SAA at a given position 
differs between the two directions.
The maps reveal two high PIN-NXB regions,
roughly coincident with small COR regions at high geomagnetic latitudes.
In addition, a slight increase ($\sim$10\%) of the PIN-NXB is 
observed in top panel just after the SAA passages.

\begin{figure}
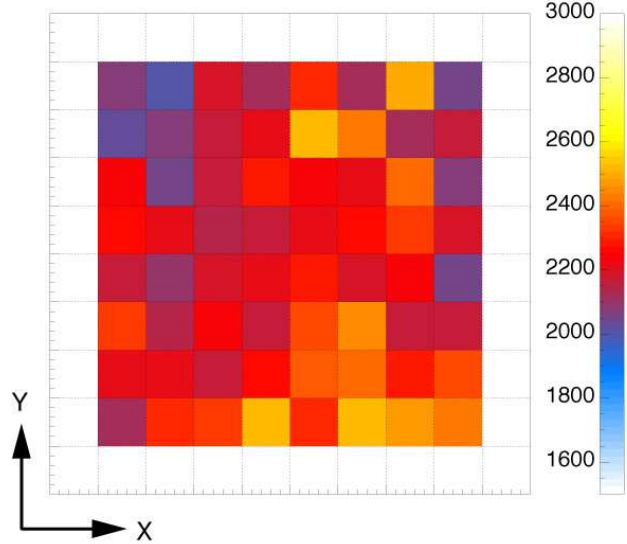

\begin{center}
\FigureFile(0.48\textwidth,0.48\textwidth){figure26.eps}
\end{center}
\caption{The distribution of integrated NXB counts from the 64 PIN diodes
with a total exposure of $\sim$2.3 Msec.
The figure is shown as the top view of HXD-S.}
\label{fig:pin_bgd_dist}
\end{figure}

In order to actually construct the NXB model as a function of geographical position,
it is necessary to accumulate the Earth occultation data at each position
using a sufficiently fine mesh covering the Earth.
During the performance verification phase,
an average exposure of the Earth occultation was $\sim$14 ks per day, 
and hence the net exposure of $\sim$3.3 Msec has already been obtained. 
This enables an NXB database to be constructed using $\sim$160 mesh points,
if requiring a minimum exposure of 20 ks each.
This  correspond to a mesh size of $\sim$30$^{\circ}\times$5$^{\circ}$ 
on geographical maps. 
If the 64 PINs have different temporal or spectral variations, 
the database should be constructed separately  for individual PINs,
and hence the modeling would becomes much more difficult.
Figure~\ref{fig:pin_bgd_dist} shows  distributions of NXB counts 
of the individual 64 PIN diodes, 
obtained in an energy range of  40--55 keV
which is free from the different LD levels. 
Deviations from the mean are confirmed to remain within $\sim$10\%, 
justifying the summed treatment of the 64 PIN diodes.

\subsection{Properties of the Residual Background of GSO}
\label{subsection:5-3}
Modeling of the NXB in GSO (hereafter GSO-NXB) is 
significantly more difficult than that of the PIN-NXB,
due to the expected presence of  {\it long-nuclides}.
Background levels due to the {\it long-nuclides}, 
to be induced in the GSO scintillator by the SAA protons, 
were estimated before  launch,
assuming that they have individually achieved equilibria in orbit.
The production cross-section for each isotope was calculated
based on a semi-empirical formula and  ground experiments (\cite{Kokubun1999}), 
and then internal activation spectra of GSO 
corresponding to decay chains of  the individual isotopes 
were constructed using detailed Monte-Carlo simulations.
However, the simulations deal only with nuclides
of which the life time is  longer than a few days, 
and hence contain neither the short-nuclides
nor the secondary emissions from cosmic-ray particles.
Therefore, detailed studies of the actual GSO-NXB are of high importance.

\begin{figure*}
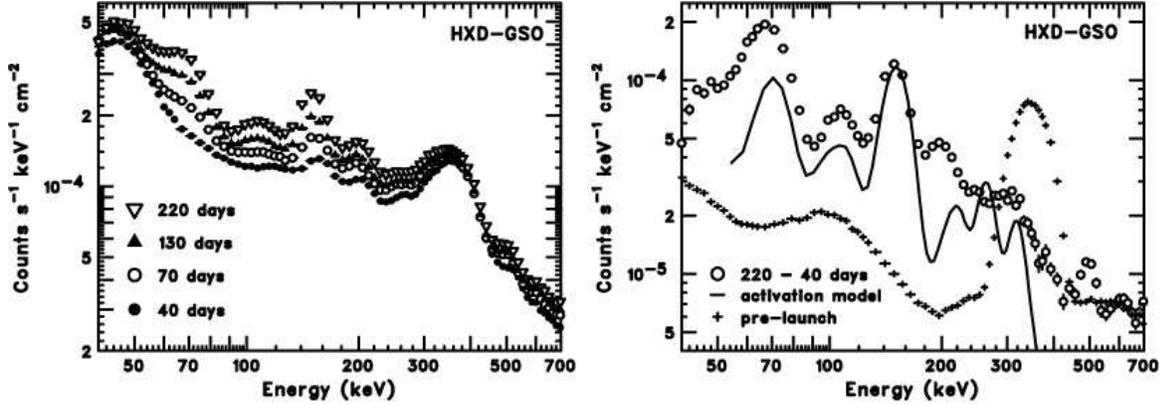

\begin{center}
\FigureFile(0.90\textwidth,0.48\textwidth){figure27.eps}
\end{center}
\caption{({\it Left}:) An evolution of averaged GSO-NXB spectra during
the first half year after the launch. 
Each observation has an exposure  longer than a day.
({\it Right}:) A difference GSO-NXB spectrum between 40 and 220 days 
after the launch (open circles), 
compared with the estimated long-nuclides spectrum (smooth curve).
The   pre-launch intrinsic background is also shown (crosses).
}
\label{fig:gso_bgd_long}
\end{figure*}

Figure~\ref{fig:gso_bgd_long} shows daily averaged GSO background spectra 
measured at 40, 70, 130, and 220 days after the launch.
The later spectra clearly show several  peaks,
which are absent or very weak in the 40-days spectrum.
These peaks, evolving on a time scale of several months,
are due to EC decays of unstable isotopes.
In contrast, the continuum  up to  400 keV  had already reached,
in the first 40 days, a relatively constant level at
$\sim 1\times 10^{-4}$ ct s$^{-1}$ keV$^{-1}$ cm$^{-2}$.
Therefore, to cancel the continuum
and extract only the ``long-nuclides'' components,
the 40-day spectrum was subtracted  from that of 220 days.
Figure~\ref{fig:gso_bgd_long} (right) compares 
this difference spectrum with the pre-launch estimation.
Thus, a good agreement is observed in positions of 
the three prominent peaks (70, 100, 150 keV), 
with a moderate accuracy within a factor of two in their absolute fluxes.
The peak at 250 keV in the pre-launch model
corresponds to \atom{Tb}{}{153}, 
and the over-estimation was probably due to 
the inadequateness of the semi-empirical formula of the nuclear 
cross-section employed in the model.
The peak at 200 keV directly originates in \atom{Eu}{}{151m}, 
an isomer with a significantly short half-life, 
but its parent nuclei is \atom{Gd}{}{151} 
whose half-life is longer than 100 days (124 d). 
From this, the three peaks have been identified
as shown in table~\ref{tbl:calline}.
The peak at 100 keV is also caused by \atom{Gd}{}{153} (150 keV peak), 
but is not listed in table~\ref{tbl:calline}, 
due to its insufficient intensity for the energy-scale calibration.

\begin{figure}
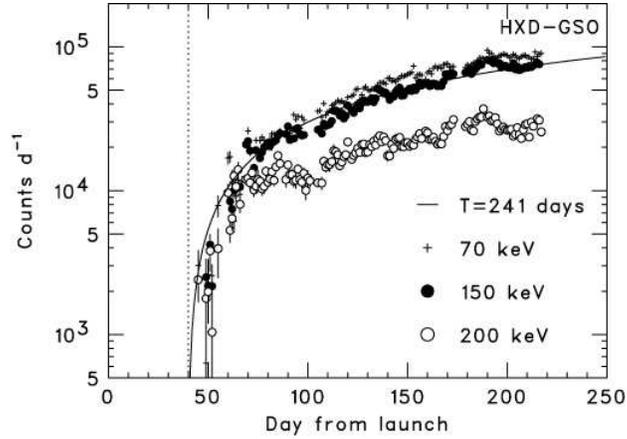

\begin{center}
\FigureFile(0.48\textwidth,0.48\textwidth){figure28.eps}
\end{center}
\caption{Light curves of the peak counts of the tree prominent EC-decay isotopes, 
compared with the predicted evolution curve for 
a half-life of 241 days. 
The difference spectra are  obtained 
by subtracting the in-orbit background taken at 40 days, 
as indicated by a dotted line.
}
\label{fig:gso_bgd_linefit}
\end{figure}

Now that individual peaks are thus identified with corresponding isotopes,
their in-orbit evolutions can be predicted from their half-lives, 
assuming that the daily dose during the SAA passages stay constant. 
Figure~\ref{fig:gso_bgd_linefit} compares this prediction
with the actually measured long-term evolution of the three prominent peaks, 
obtained by fitting the difference spectra 
(with the 40-days spectrum subtracted)
with multiple gaussian functions.
The light curve of the 150 keV ($^{151}$Gd) peak
indeed shows a good agreement with a calculation
assuming a half-life of 241 days,
hence confirming the isotope identification.
Since the longest life of the identified isotopes is shorter than a year, 
the long-term increase in the GSO-NXB is expected to saturate
on a time scale of a year.
Then, it will show a modulation within a factor of two, 
anti-correlated with the solar activitiy cycle.

\begin{figure}
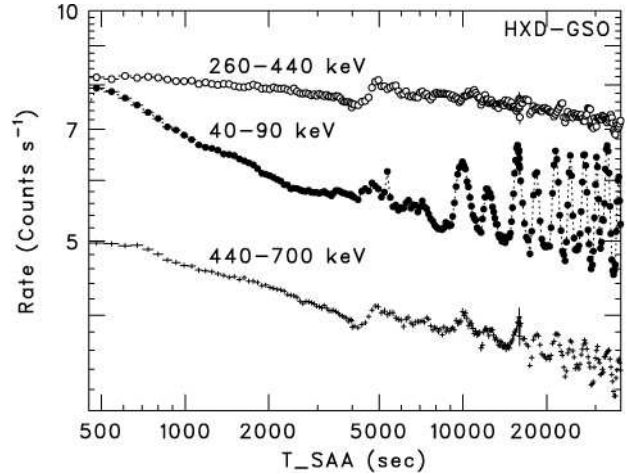

\begin{center}
\FigureFile(0.48\textwidth,0.48\textwidth){figure29.eps}
\end{center}
\caption{Light curves of the GSO-NXB in 40--90, 260--440, and
440--700 keV, folded with the T\_SAA.
}
\label{fig:gso_bgd_lc}
\end{figure}

Figure \ref{fig:gso_bgd_lc} shows the GSO-NXB counts 
in several broad energy bands,
folded with the T\_SAA parameter (\S\ref{subsection:5-2}).
The light curves in the lowest and highest energy bands
thus decline rapidly  with  T\_SAA,
suggesting the dominance of short-nuclides
in the lowest end of the GSO energy range 
and at around the annihilation line energy.
In contrast,  the background rate near the intrinsic \atom{Gd}{}{152}
peak (Paper I) is almost independent of T\_SAA.
In  non-SAA orbits which correspond to T\_SAA values longer than 5000 s, 
the GSO-NXB is seen to anti-correlate with  COR,
in all  energy bands but with varying amplitudes,
with a maximum of $\sim$40\%. 
These results imply that the spectral shape of  GSO-NXB 
depends  on both T\_SAA and COR.

\begin{figure*}
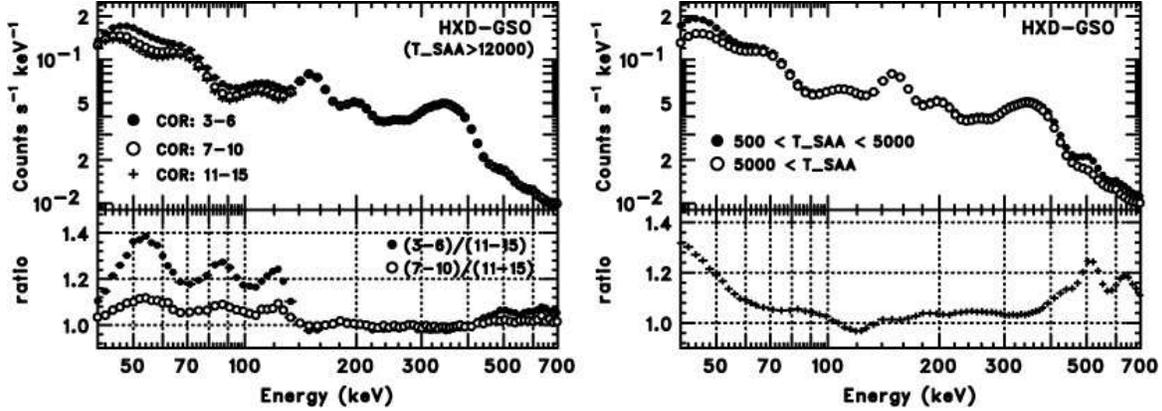

\begin{center}
\FigureFile(0.90\textwidth,0.48\textwidth){figure30.eps}
\end{center}
\caption{
({\it Left:}) The GSO background spectra accumulated at
several COR regions of non-SAA orbits.
({\it Right}:) The GSO background spectra accumulated 
from SAA passages (filled circle) and non-SAA orbits (open circle). 
}
\label{fig:gso_bgd_spec}
\end{figure*}

The GSO-NXB spectra, sorted by COR and T\_SAA, 
are shown in the left and right panels of figure \ref{fig:gso_bgd_spec}, respectively. 
As expected from figure \ref{fig:gso_bgd_lc}, 
large variations are  found mostly below 150 keV,
while the spectral shape stays rather constant at 150--400 keV. 
The annihilation line, which is probably caused by 
$\beta^{+}$-decays in the surrounding BGO scintillators or passive materials,
becomes obvious when the spectrum during SAA orbits is 
compared with that during non-SAA environment,
indicating a rapid progress of $\beta^{+}$-decay processes.
The ratio spectrum between small and large COR regions
shows three prominent peaks, 
one of which corresponds to the K-edge energy of gadolinium ($\sim$50 keV).
Since  prompt emission from the cosmic-ray particle events are
already discarded by the hard-wired PSD and mutual anti-coincidence, 
these peaks suggest the  existence of significant short-nuclides
induced by the primary events.

\subsection{Comparison with Other Missions}
\label{subsection:5-4}

\begin{figure}
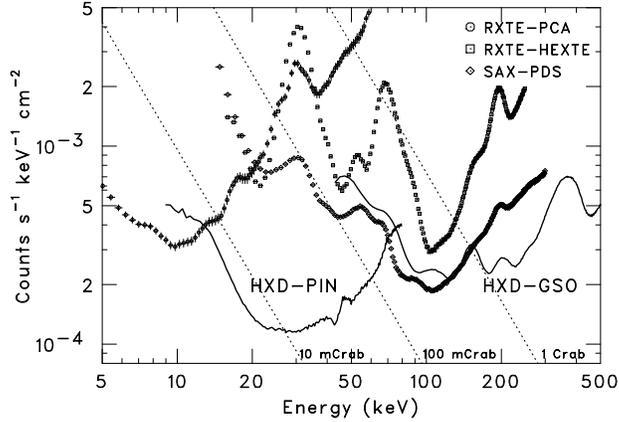

\begin{center}
\FigureFile(0.48\textwidth,0.48\textwidth){figure31.eps}
\end{center}
\caption{The in-orbit detector background of PIN/GSO,
averaged over 2005 August to 2006 March and 
normalized by individual effective areas. 
For comparison, those of the  RXTE-PCA, RXTE-HEXTE, and BeppoSAX-PDS
are also shown.
Dotted lines indicate 1 Crab, 100 mCrab, and 10 mCrab intensities.
}
\label{fig:nxb_compare}
\end{figure}

As described in \S\ref{subsection:5-1}--\S\ref{subsection:5-3},
the temporal and spectral behavior of the PIN-NXB and GSO-NXB 
can be both described basically in terms of the satellite position in orbit.
Any unexpected or sporadic variations are insignificant, 
which in turn ensures, at least potentially,
an accurate background modeling. 
Especially, the absence of long-term changes in the PIN-NXB
is quite advantageous in constructing an accreate model. 
In figure \ref{fig:nxb_compare}, 
the NXB spectra of PIN and GSO,
averaged over the performance verification phase, 
are compared with typical in-orbit detector backgrounds 
of other non-imaging hard X-ray detectors.
In energy ranges of 15--70 and 150--500 keV, 
the lowest background level has been achieved by the HXD. 
Since the averaged HXD spectra include both SAA and non-SAA orbits, 
the HXD backgrounds can be reduced by 2--20\%,
especially in the GSO background below 100 keV,
if data from only non-SAA orbits are employed.

While the sensitivity of rocking detectors are limited
by statistical errors in off-source and on-source spectra,
that of the HXD is solely determined by the accuracy 
of background modeling, 
in the case of background dominant sources.
Two types of PIN-NXB models have been developed and tried so far.
The first one estimates the background flux at a given time
based on  the instantaneous PIN-UD rate (summed over the 64 PINs) at that time 
and the PIN-UD rate time-integrated with a certain decay time constant,
while the other employs a model fitting to 
the PIN-UD light curve from each  observation. 
Results from the two independent methods 
have been confirmed to agree within an  accuracy of 5\%, 
and both have been already applied to scientific analyses
of celestial sources with an intensity of a few tens percent of the background.
Except the SAA orbits in which the PIN-NXB increases significantly,
the reproducibility of both models have been confirmed to be 3--5\%.
Construction of more accurate models, 
which are applicable to the SAA orbits with an accuracy better than 3\%,
are  in progress, 
and results will be made available to the public.

%
%
\section{Other Calibration Items}
\label{section:6}

\subsection{Angular Response}
\label{subsection:6-1}

\begin{figure}
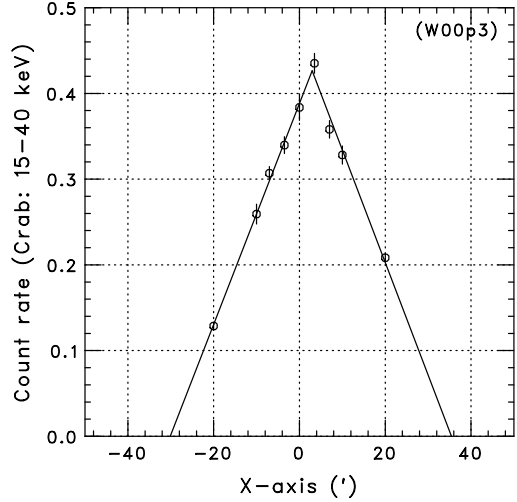

\begin{center}
\FigureFile(0.40\textwidth,0.40\textwidth){figure32.eps}
\end{center}
\caption{Typical angular response of a single fine collimator 
along the satellite X-axis, 
obtained from the nine offset observations of the Crab nebula.}
\label{fig:fcalign_x}
\end{figure}

Individual fields-of-view of the 64 PIN diodes are 
collimated with 64 passive fine collimators to $\sim$34$'$. 
Although it was confirmed in the on-ground calibrations 
that the optical axes of all fine collimators are  aligned 
within an accuracy of 3.5$'$ (Paper I),
in-orbit measurements are inevitable to investigate the launch vibration effect.
In addition, the absolute alignment of the HXD optical axis
to the spacecraft and the XRT/XIS system must be reconfirmed in orbit.
Multiple pointing observations on the Crab nebula were thus performed,
which consisted of 0$'$, 3.5$'$, 7$'$, 10$'$, and 20$'$ offset positions in 
both $\pm$X and $\pm$Y directions (\cite{Serlemitsos2006}).
Background extracted spectra were constructed 
for all the 64 PIN diodes individually, from every observation, 
and counting rates at that location were obtained 
in an energy range of 15--40 keV after the dead time correction
(\S\ref{subsection:6-2}).
These fluxes were plotted against X- and Y-axis of the spacecraft
coordinates, and the X/Y central axis of the collimator was 
calculated by fitting the flux distribution with a trianglar function.
An example of X-axis angular response is shown in figure \ref{fig:fcalign_x}.
The fitting procedure was repeated, 
excluding data points which are closer than 3.5$'$ from the center
derived from the first trial, 
since the Crab nebula is not exactly a point source.

\begin{figure}
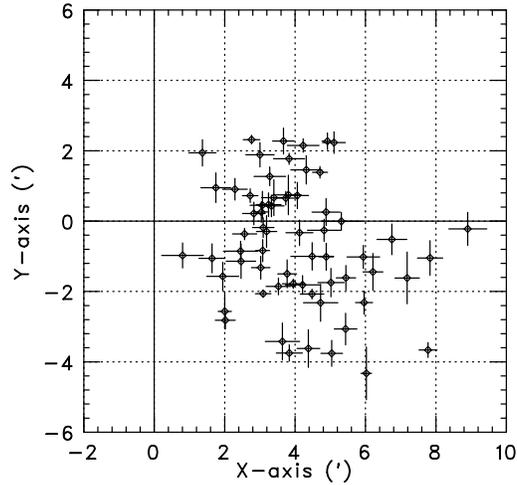

\begin{center}
\FigureFile(0.40\textwidth,0.40\textwidth){figure33.eps}
\end{center}
\caption{A summary plot of the distribution of individual optical
axes of the 64 fine collimators.}
\label{fig:fcalign_xy}
\end{figure}

\begin{figure}
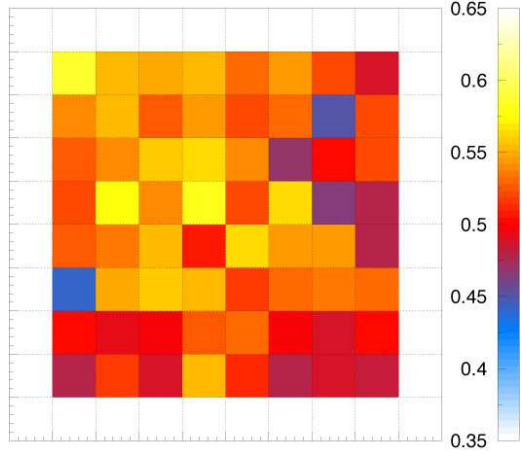

\begin{center}
\FigureFile(0.40\textwidth,0.40\textwidth){figure34.eps}
\end{center}
\caption{
The distribution of counting rates of the Crab nebula,
obtained from the 64 PIN diodes in an energy range of 15--40 keV.
The observation was performed at the HXD nominal position.
The figure illustrates a top view of HXD-S.
}
\label{fig:crab_ratemap}
\end{figure}

Figure \ref{fig:fcalign_xy} shows the resultant distributions
of 64 optical axes of the fine-collimators in the X and Y directions.
The alignment of each collimator was determined with a typical error of $\sim$1$'$,
and the overall  scatter among the 64 was confirmed  to remain within $3'.5$ (FWHM)
from the average.
However, the weighted mean shows a slight offset
by $\sim4'$ in the X-direction from the optical axis of Suzaku
(i.e., the XIS-nominal position).
This offset brings a typical decrease by $\sim$10\% in the effective area 
when an observation is performed at the XIS-nominal position.
As shown in figure \ref{fig:crab_ratemap}, 
even at the HXD-nominal position, individual effective areas of
the 64 PIN diodes vary at $\sim$10\% level, 
and hence this effect is taken into account in the energy response matrix.

\subsection{Dead Time}
\label{subsection:6-2}

\begin{figure}
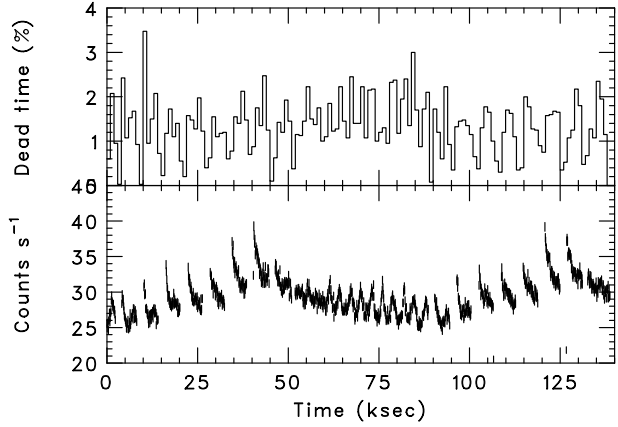

\begin{center}
\FigureFile(0.48\textwidth,0.48\textwidth){figure35.eps}
\end{center}
\caption{A typical light curve of the dead time fraction of HXD ({\it top}),
and total event rate ({\it bottom}), measured during 1.5 days in orbit.
Since the event output is disabled during the SAA passages, 
the actual dead time is 100\% in that period.
}
\label{fig:dtlc}
\end{figure}

The HXD dead time, 
contained after all the screening procedures have been applied,
is determined by following three factors: dead time caused
by the hard-wired electronics in HXD-AE (Paper I), 
that due to the limitation of data transfer rate between HXD-AE and HXD-DE,
and that due to the telemetry saturation.
The third case is usually avoided by adequately setting 
the onboard hardware and software (\S\ref{subsection:2-4}), 
and even if there is any period of the telemetry saturation,
that interval will be eliminated by the offline analysis software.
To accurately estimate the first and second components,
``pseudo events'' are regularly triggered in HXD-AE
with a period of an event per four seconds per one Well unit (Paper I).
Since the pseudo events are discarded 
if the pseudo trigger is generated 
while a ``real event'' is inhibiting other triggers, 
the dead-time fraction can be estimated by counting 
a number of pseudo events, output to the telemetry,
and comparing with the expected counts during the same exposure.
Figure \ref{fig:dtlc} shows a typical light curve of thus estimated
dead time fraction. 
Since the event output, except for the pseudo events, 
is disabled during the SAA passages, 
the fraction drops down to nearly zero in every SAA,
while it reaches at most $\sim$3\% when the event trigger rate 
becomes high due to the activation. 
These values are consistent with a rough estimation; 
a multiplication of the average of trigger rate ($\sim$1000 ct s$^{-1}$)
and a typical duration of data acquisition sequence ($\sim$25$\mu$s).

In addition to the dead time, events can be randomly discarded 
by both the chance coincidence in the hit-pattern flags 
and PSD selection of GSO. 
While the latter probability is counted in the energy response
matrices of GSO, based on the width of selection criteria 
(\S\ref{subsection:4-4}),
the former can be estimated again using the pseudo events.
Since the hit-pattern signals are latched when the pseudo trigger
has been generated, in the same manner as the real trigger, 
the chance probability is derived by applying the same anti-coincidence
conditions to the pseudo events as those utilized for the true events,
and count the number of discarded ones. 
The chance coincidence estimated by this method 
further reduces an ``effective exposure'' by 3--5\% fraction,
which is also consistent with that expected from the width of 
hit-pattern signal (5.6 $\mu$s) and averaged counting rate of 
the SLD ($\sim$1000 ct s$^{-1}$).
Since the SLD rate from Well unit is dominated by the
activation of BGO scintillators, 
the chance coincidence probability is hardly affected 
by source intensities, 
which is less than 1\% of the average rate even in case of 
the Crab nebula ($\sim$6 ct s$^{-1}$ per Well).

\begin{figure}
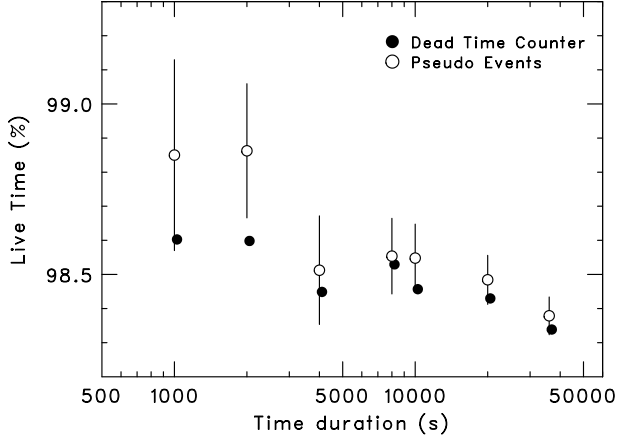

\begin{center}
\FigureFile(0.48\textwidth,0.48\textwidth){figure36.eps}
\end{center}
\caption{Comparison of dead times, 
calculated with the pseudo event and the dead time counter.}
\label{fig:dtcorr}
\end{figure}

The dead time counter in HXD-AE (Paper I) can be used as 
another method to estimate the onboard dead time.
While the estimation utilizing the pseudo events suffers from
a large statistical error in case of a short exposure,
the dead time counter uses 156 kHz clock as a base, 
and hence can accurately estimate even with a short duration.
Figure \ref{fig:dtcorr} shows a comparison between the two 
methods at several exposure times. They show an agreement
within the statistical error.

\subsection{Timing Accuracy}
\label{subsection:6-3}

The mission requirements on timing is
100 $\mu$sec to the relative and absolute timing,
and $10^{-8}$ order of the stability.
The arrival time of events detected by the HXD is designed
to have a resolution of 61 $\mu$sec in the "normal" mode, and 31 $\mu$sec
in the "fine" mode.
Since only 19 bits per event are reserved for the timing information
in the limited word size (16 bytes) of the event data,
there is a very large gap in size between the raw data (19 bits) and 
the mission time records of over 10 years life 
($10^{15} \times$ 100 $\mu$sec).
The timing system of the HXD is designed to connect with three-types of
timing counters;
19 bits event counter with 61 $\mu$sec to 32 second coverage,
timing counter in the central data processing unit (DP) of the satellite
with $1/4096$ seconds to 1 M seconds, and the ground Cesium clocks at 
the ground station.
All of the timing counters in the {\it Suzaku} satellite are originated
from only one crystal oscillator in DP which has a timing stability of
about $2 \times 10^{8}$ order after correcting drifts
by variable temperature of the oscillator.
The absolute timing is re-calibrated at every time of a contact passage
to the ground station, which appears 5 revolutions par day
with about 10 minutes duration for each.

\begin{figure}
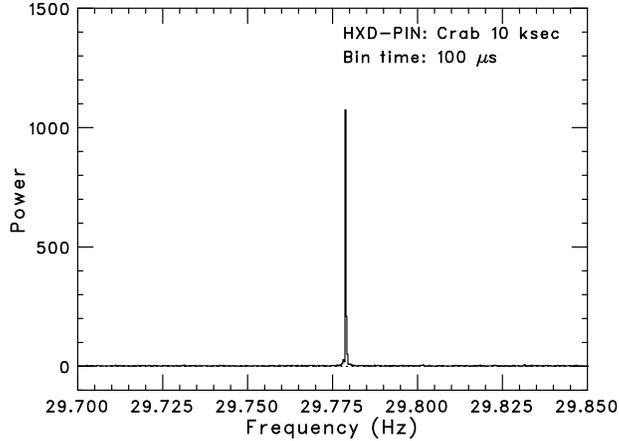

\begin{center}
\FigureFile(0.48\textwidth,0.48\textwidth){figure37.eps}
\end{center}
\caption{A power spectrum of the Crab pulsar after the barycentric correction, 
obtained as a sum of the 64 PIN diodes.
}
\label{fig:crab_power}
\end{figure}

\begin{figure}
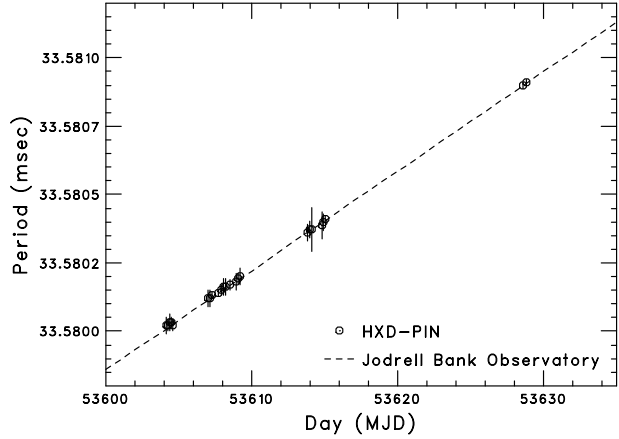

\begin{center}
\FigureFile(0.48\textwidth,0.48\textwidth){figure38.eps}
\end{center}
\caption{
Pulse periods of the Crab pulsar obtained in the 15--40 keV band with PIN, 
from 24 observations performed in 2005 August and September.
The dashed line shows those measured in the radio wavelength.
}
\label{fig:crab_period}
\end{figure}

\begin{figure}
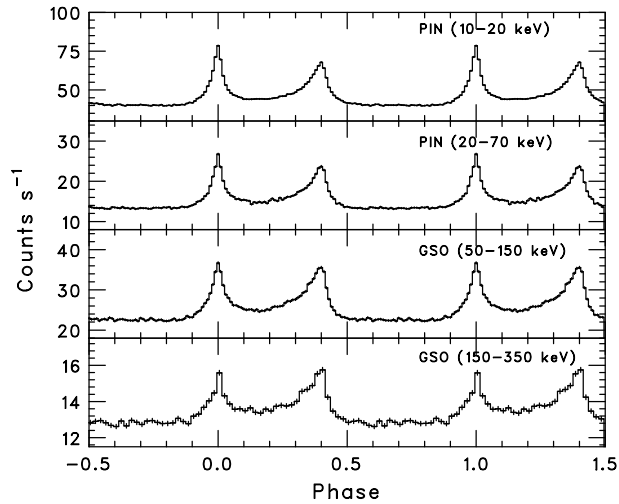

\begin{center}
\FigureFile(0.48\textwidth,0.48\textwidth){figure39.eps}
\end{center}
\caption{
Folded light curves of the Crab pulsar in four energy bands,
obtained as a sum of the 64 PIN diodes  or the 16 Well units.
The observed counts are folded at the pulse period, and are shown
for two pulse phases.
}
\label{fig:crab_folded_lc}
\end{figure}

In-orbit timing calibrations in the initial phase are performed 
by X-ray pulsars or binaries,
such as the Crab pulsar, PSR1509$-$58, and Hercuris X-1,
with 33 millisecond, 150 millisecond and $\sim$1.0 second period, 
respectively.
As shown in figure~\ref{fig:crab_power},
clear pulsations are detected from all of the Crab observations,
after applying the barycentric correction.
The pulse periods derived by the HXD (33.58087$\pm$0.00001 ms) 
show a good agreement with that obtained from a simultaneous 
radio observation (33.5808764 ms).
As shown in figure~\ref{fig:crab_period},
the relative timing stability is thus confirmed 
as $10^{8}$ order by a series of simultaneous Crab observations.
The pulse profiles obtained at several energy bands, 
as shown in figure~\ref{fig:crab_folded_lc}, 
are also confirmed to be consistent with those obtained in the similar band
with other gamma-ray missions; 
RXTE \citep{Rots2004} and INTEGRAL \citep{Mineo2006}. 
Besides the hybrid detection devices inside the Well unit,
the HXD can also detect Gamma-ray bursts (GRBs) by the Anti units
with a timing resolution of 15 (or 31) $\mu$sec (Paper I). 
The absolute timings are recorded by TPU modules in HXD-AE, 
and used for the Inter Planetary Network system.
The timing accuracy for GRB triggers was confirmed to 
be consistent with those by other $\gamma$-ray missions, like
{\it Swift}, {\it Konus-Wind}, {\it HETE-2} and {\it INTEGRAL}
in 2 msec \citep{Yamaoka2006}.

\subsection{Cross-calibration with XIS}
\label{subsection:6-4}

\begin{figure*}
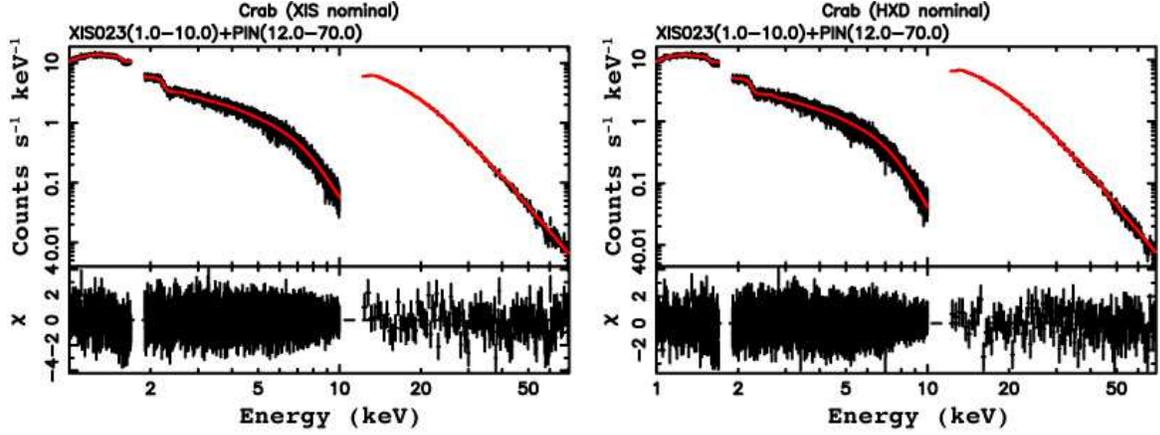

\begin{center}
\FigureFile(0.90\textwidth,0.48\textwidth){figure40.eps}
\end{center}
\caption{The background-subtracted Crab spectrum of XIS
(summed over the three XIS-FIs) and PIN,
compared with the best fit absorbed power-law model.
An energy range of 1.7--1.9 keV is excluded from the XIS spectra,
due to large systematic uncertainties of the current response matrices
({\tt ae\_xi[023]\_20060213.rmf} and 
{\tt ae\_xi[023]\_[xis/hxd]nom6\_20060615.arf}).
}
\label{fig:suzaku_crab}
\end{figure*}

As described in \citet{Koyama2006} and \citet{Serlemitsos2006},
in-orbit calibrations of XISs and XRTs have been extensively
performed in parallel with those of the HXD, 
to realize the wide-band spectroscopy with Suzaku. 
Until now, 
two instruments have been independently calibrated,
and no ``adjustment'' has been yet performed.
Therefore, 
even though the XIS and HXD always simultaneously observe the Crab nebula,
individual spectral fittings result
slight differences between individual best fit parameters,
on both the photon index and absolute flux
at a level of $\sim$0.04 and $\sim$15\%, respectively.
As shown in figure \ref{fig:suzaku_crab}, 
the overall spectra can still be reproduced well
over a wide energy range of 1--70 keV,
when fitted with power-law models having a common 
photon index and different normalizations between the XIS and PIN.
When a ``constant factor'', 
which represents the relative normalization of PIN 
above the average of three front illuminated CCD cameras (XIS-FIs),
is introduced as shown in table \ref{tbl:suzaku_crabfit},
simultaneous fittings with a single power-law model gives
higher absolute fluxes from PIN diodes than those of the averaged XIS-FIs
with a level of $\sim$13\% and $\sim$15\%, 
at the XIS and HXD nominal positions, respectively.
This overestimation infers that thinner depletion layers 
than true thicknesses were employed in the mass model 
(\S\ref{subsection:3-4}), 
and remains as an issue to be further investigated.

\begin{longtable}{lcccc}
\caption{Best-fit parameters and 90\% confidence errors for the 
spectra of the Crab Nebula at the XIS and HXD nominal positions.}
\label{tbl:suzaku_crabfit}
\hline\hline
 Position & $N_{\rm H}$\footnotemark[$*$] & 
 Photon index & Normalization\footnotemark[$\dagger$] & 
Constant factor\footnotemark[$\ddagger$]\\
\endfirsthead
\hline\hline
 Position & $N_{\rm H}$\footnotemark[$*$] & 
 Photon index & Normalization\footnotemark[$\dagger$] & 
Constant factor\footnotemark[$\ddagger$]\\
\endhead
\hline
\endfoot
\endlastfoot
\hline
XIS nominal\footnotemark[$\S$] & 
0.32 $\pm$ 0.01 & 2.10 $\pm$ 0.01 & 10.0 $\pm$ 0.1 & 1.13 $\pm$ 0.01 $\pm$ 0.02 \\
HXD nominal\footnotemark[$\|$] & 
0.30 $\pm$ 0.01 & 2.09 $\pm$ 0.01 &  9.5 $\pm$ 0.1 & 1.15 $\pm$ 0.01 $\pm$ 0.02 \\
\hline
\multicolumn{5}{l}{\hbox to 0pt{\parbox{180mm}{\footnotesize
\par\noindent
\footnotemark[$*$] Hydrogen column density in a unit of 10$^{22}$ cm$^{-2}$.
\par\noindent
\footnotemark[$\dagger$] Power-law normalization in a unit of 
photons cm$^{-2}$ s$^{-1}$ keV$^{-1}$ at 1 keV.
\par\noindent
\footnotemark[$\ddagger$] Relative normalization of PIN above XIS.
\par\noindent
\footnotemark[$\S$] Observation performed on 2005 Sep.15 19:50--Sep.16 02:10 (UT)
\par\noindent
\footnotemark[$\|$] Observation performed on 2005 Sep.15 14:00--19:50 (UT)  
}\hss}}
\end{longtable}

%
%
\section{Summary and Conclusion}
\label{section:7}

The results of in-orbit performances and calibrations 
of the HXD can be summarized by the following points:\\

\begin{enumerate}
\item
The initial run-up operations of the HXD and fine tunings of 
HXD-AE and HXD-DE were completed at 40 days after the launch of Suzaku.
The instrument was confirmed to have survived 
the launch vibrations and controlled rapid decrease of temperature
with no significant damage. 
\item
The nominal in-orbit operation mode, which includes 
the high-voltage levels for PIN diodes and PMTs, 
fine gain settings, lower and upper threshold levels,
and PSD selection conditions for scintillator events
were basically established on 2005 August 19, 
and only slightly changed during the performance
verification phase. 
The onboard background reduction system
based on the anti-coincidence method 
was confirmed to function effectively.
\item
The in-orbit energy scale of every PIN diode was confirmed
to be quite stable, 
and was accurately determined with an accuracy of 1\%.
Individual lower energy thresholds ranging 9--14 keV 
were successfully adopted to the 64 PINs.
\item
The in-orbit energy scales of the GSO scintillators were
determined in view of the temporal variations. 
Below 100 keV, they showed the additional nonlinearities.
\item
The event selection conditions utilized in the analysis software
were optimized in terms of the signal acceptance and chance 
coincidence. 
The residual in-orbit PIN-NXB level was confirmed to be as low as
$\sim$0.5 ct s$^{-1}$, 
which corresponds to about 10 mCrab intensity.
\item
The temporal and spectral behaviors of the PIN-NXB and GSO-NXB 
were extensively studied. 
They show individual dependencies mainly on the cut-off rigidity
and elapsed time after the SAA passage, 
in addition to the long-term accumulation of in-orbit activations.
\item
The NXB modeling is still in progress.
The current uncertainty of PIN-NXB models are $\sim$5\%.
\item 
The energy response matrices of PIN and GSO were constructed,
and confirmed to reproduce the Crab spectrum at 12--70 and 100-300 keV
energy range, 
with typical accuracies of $\sim$5\% and $\sim$10\%, respectively.
\item
The individual alignment of the 64 fine-collimators were determined
in orbit with a typical error of $\sim$1$'$.
\item
The instrumental dead-time was confirmed to be 1--2\% level, 
while the chance coincidence probability further reduces 3--5\%
of the effective exposure.
\item
The HXD timing accuracy was confirmed to be normal.
\item
The relative normalization of PIN above the XIS was derived as 13--15\% level, 
at the current calibration status.

\end{enumerate}

\bigskip

The authors are deeply grateful for dedicated contributions 
provided by the former members of the development team.
This research was partially supported by the Ministry of Education, 
Science, Sports and Culture, Grant-in-Aid for Scientific Research 
for Priority Areas, 14079201, 2006.


\end{document}